\documentclass[aps,prl,twocolumn,superscriptaddress,preprintnumbers]{revtex4-1}
\usepackage{lipsum}
\usepackage{amsmath}	%Allows \begin{equation}, \begin{align}, \underbrace{2+2=4}_{\text{obvious}}
\usepackage{bm}		%Allows {\bm \nabla \alpha} for bold symbold
\usepackage{amssymb}	%Allows use math symbols
\usepackage{color} 		%Allows {\color{red} Hello World} or \colorbox{blue}{Hello World}
\usepackage{graphicx} 	%Allows \includegraphics{figure.pdf}
\usepackage[T1]{fontenc}
\usepackage{times} 		%Use Times New Roman font
\usepackage{mathrsfs} %Allows \mathscr{HELLO}
\usepackage{hyperref} %Automatically links \label and \ref commands; Always load last
\usepackage[utf8]{inputenc}
\usepackage{multirow, booktabs}
\hypersetup{
    colorlinks=true,       % false: boxed links; true: colored links
    linkcolor=blue,        % color of internal links
    citecolor=blue,        % color of links to bibliography
    filecolor=magenta,     % color of file links
    urlcolor=blue          % color of external links
}
\usepackage[all]{hypcap} %Link navagates to top of figure instead of caption (below fig)

\newcommand{\es}[2] {\begin{equation} \label{#1} \begin{split} #2 \end{split} \end{equation}}

\newcommand{\be}{\begin{equation}}
\newcommand{\ee}{\end{equation}}
\newcommand{\bea}{\begin{eqnarray}}
\newcommand{\eea}{\end{eqnarray}}

\newcommand{\gsim}{\lower.7ex\hbox{$\;\stackrel{\textstyle>}{\sim}\;$}}
\newcommand{\lsim}{\lower.7ex\hbox{$\;\stackrel{\textstyle<}{\sim}\;$}}

\usepackage{array}

\newcommand{\thickhline}{%
    \noalign {\ifnum 0=`}\fi \hrule height 1pt
    \futurelet \reserved@a \@xhline
}
\newcolumntype{"}{@{\hskip\tabcolsep\vrule width 1pt\hskip\tabcolsep}}
\makeatother

\newcommand{\new}[1]{\textcolor{black}{#1}}

\begin{document}

\title{\new{Axion Emission Can Explain a New Hard $X$-ray Excess from Nearby Isolated Neutron Stars}}
\author{Malte Buschmann}
\affiliation{Leinweber Center for Theoretical Physics, University of Michigan, Ann Arbor, MI 48109 U.S.A.}\affiliation{Department of Physics, Princeton University, Princeton, NJ 08544}

\author{Raymond T. Co}
\affiliation{Leinweber Center for Theoretical Physics, University of Michigan, Ann Arbor, MI 48109 U.S.A.}
\affiliation{William I. Fine Theoretical Physics Institute, School of Physics and Astronomy, University of Minnesota, Minneapolis, MN 55455, USA}

\author{Christopher Dessert}
\affiliation{Leinweber Center for Theoretical Physics, University of Michigan, Ann Arbor, MI 48109 U.S.A.}
\affiliation{Berkeley Center for Theoretical Physics, University of California, Berkeley, CA 94720, U.S.A.}
\affiliation{Theoretical Physics Group, Lawrence Berkeley National Laboratory, Berkeley, CA 94720, U.S.A.}

\author{Benjamin R. Safdi}
\affiliation{Leinweber Center for Theoretical Physics, University of Michigan, Ann Arbor, MI 48109 U.S.A.}
\affiliation{Berkeley Center for Theoretical Physics, University of California, Berkeley, CA 94720, U.S.A.}
\affiliation{Theoretical Physics Group, Lawrence Berkeley National Laboratory, Berkeley, CA 94720, U.S.A.}

\date{\today}

\begin{abstract}
\new{
Axions may be produced thermally inside the cores of neutron stars (NSs), escape the stars due to their feeble interactions with matter, and subsequently convert into $X$-rays in the magnetic fields surrounding the stars.  We show that a recently-discovered excess of hard $X$-ray emission in the 2 - 8 keV energy range from the nearby Magnificent Seven isolated NSs could be explained by this emission mechanism.  These NSs are unique in that they had previously been expected to only produce observable flux in the UV and soft $X$-ray bands from thermal surface emission at temperatures $\sim$100 eV.  No conventional astrophysical explanation of the Magnificent Seven hard $X$-ray excess exists at present.  We show that the hard $X$-ray excess  may be consistently explained by an axion-like particle with mass $m_a \lesssim 2 \times 10^{-5}$ eV 
and $g_{a\gamma\gamma} \times g_{ann} \in (2 \times 10^{-21}, 10^{-18})$ GeV$^{-1}$ at 95\% confidence, accounting for both statistical and theoretical uncertainties, where $g_{a\gamma\gamma}$ ($g_{ann}$) is the axion-photon (axion-neutron) coupling constant. 
}
\end{abstract}

\preprint{LCTP-19-26}

\maketitle

Neutron stars (NSs) have long been recognized as excellent laboratories for searching for new light and weakly coupled particles of nature.  This is because such particles may be produced abundantly in the hot cores of the NSs, escape, and thus provide a pathway by which the NSs may cool.  Some of the strongest constraints on the ultralight pseudo-scalar particles known as axions arise from NS cooling~\cite{Umeda:1997da,Keller:2012yr,Leinson:2014ioa,Sedrakian:2015krq,Sedrakian:2018kdm}. 
{  Axions may be produced through nucleon bremsstrahlung in various combinations of proton and neutron scattering in the NS cores~\cite{Iwamoto:1984ir,Brinkmann:1988vi}.} 
    It has also been suggested the axions produced in the NS cores may convert into $X$-rays in the magnetospheres surrounding the NSs and that these $X$-rays may be observable~\cite{Morris:1984iz,Raffelt:1987im,Fortin:2018ehg}.        

\new{
In this work we provide a consistent interpretation of the recently-observed hard $X$-ray excess from the nearby Magnificent Seven (M7) $X$-ray dim isolated NSs~\cite{dessert2019hard} in the context of an axion model.
Ref.~\cite{dessert2019hard} found significant excesses of hard $X$-ray emission, in the 2 - 8 keV energy range, from the M7 using data from the {\it XMM-Newton} and {\it Chandra} $X$-ray telescopes.
In particular,~\cite{dessert2019hard} found that the NS RX J1856.6-3754 (J1856) has around a 5$\sigma$ excess, RX J0420.0-5022 (J0420) has a $\sim$3$\sigma$ excess, and RX J1308.6+2127 (J1308) has a $\sim$2$\sigma$ excess.
The NSs RX J2143.0+0654 (J2143) and RX J0720.4-3125 (J0720) have marginal $\sim$1$\sigma$ excesses, while RX J1605.3+3249 (J1605) has a small deficit and is consistent with zero hard $X$-ray flux. 
We show that the M7 hard $X$-ray data may be explained in the context of an axion model where the axion couples to both nucleons and photons.  The fact that hard $X$-ray emission is observed from some NSs and not others is consistent with the axion model because (i) the exposure times vary across the M7, (ii) the predicted fluxes at fixed axion parameters vary between NSs, given their different properties, and (iii) these properties are uncertain at present.  We also provide one of the most competitive constraints to-date on the axion-photon times axion-nucleon coupling for axion masses $m_a \lesssim 10^{-4}$ eV.
}

The M7 were discovered in soft $X$-rays with the {\it ROSAT} All Sky Survey (see, {\it e.g.},~\cite{Haberl2007}).  Their soft spectra are well described by near-thermal distributions with surface temperatures $\sim$50-100 eV.  No non-thermal emission, for example in radio, has been previously observed from the NSs.  As such, they are expected to produce negligible hard $X$-ray flux, making them background-free from the point of view of the analysis described in this work.  Moreover, they are all observed to have strong magnetic fields~\cite{Kaplan:2005fy,Hambaryan:2017wvm,Kaplan:2005zr,vanKerkwijk:2007jp,Kaplan:2009au,Kaplan:2009ce,Kaplan:2011xd,Malacaria:2019zqr} 
and to be relatively nearby, at distances of order hundreds of pc.     

The quantum chromodynamics (QCD) axion is a hypothetical ultralight particle that solves the strong {\it CP} problem of the neutron electric dipole moment~\cite{Peccei:1977ur,Peccei:1977hh,Weinberg:1977ma,Wilczek:1977pj} and may also make up the observed dark matter~\cite{Preskill:1982cy,Abbott:1982af,Dine:1982ah}. 
{The QCD axion and axion-like particles (ALPs) more generally also appear to be a relatively generic expectation from string compactifications~\cite{Svrcek:2006yi,Arvanitaki:2009fg}, and the ALP masses in particular may be significantly lighter than the $\sim$$10^{-4}$ eV threshold relevant for this work (see, {\it e.g.},~\cite{Hui:2016ltb,Stott:2018opm}). }
 Both the QCD axion and ALPs are expected to couple derivatively to matter and also couple to electromagnetism, allowing them to be produced inside of the hot NSs and converted into photons in the strong magnetic fields surrounding the NSs.  Thus in this work we refer to both particles simply as axions.
    Intriguingly, recent string theory constructions suggest that the ALP photon couplings may be slightly smaller than current limits and within reach of the search discussed in this work~\cite{Halverson:2019cmy}.

Axions have also been discussed in the context of white dwarf, red giant, and horizontal-branch (HB) star cooling~\cite{Raffelt:1985nj,Isern:2008nt,Isern:2008fs,Isern:2010wz,Bertolami:2014wua,Ayala:2014pea,Redondo:2013wwa,Viaux:2013lha,Giannotti:2015kwo,Giannotti:2017hny}.  In white dwarf (WD) and red giant stars the dominant production modes involve the axion-electron coupling while in HB stars the axion-photon production dominates.  Recently it was proposed that $X$-ray observations of magnetic WD stars may probe axion scenarios, since the hot axions produced in the WD cores may convert into $X$-rays in the magnetic fields surrounding the WDs~\cite{Dessert:2019sgw}. 
Axion-photon conversion within NS magnetospheres has been discussed recently in the context of dark matter axions~\cite{Pshirkov:2007st,Hook:2018iia,Safdi:2018oeu,Edwards:2019tzf}.   
 Axions and axion dark matter are also the subject of considerable laboratory searches~\cite{Shokair:2014rna,Du:2018uak,Brubaker:2016ktl,Kenany:2016tta,Brubaker:2017rna,TheMADMAXWorkingGroup:2016hpc,Kahn:2016aff,Foster:2017hbq,Ouellet:2018beu,Chaudhuri:2014dla,Silva-Feaver:2016qhh,Budker:2013hfa,Bahre:2013ywa,Bogorad:2019pbu,Janish:2019dpr,Lawson:2019brd}.

\new{
This work takes the M7 hard $X$-ray spectra from~\cite{dessert2019hard} as a starting point.
Additional analysis details and systematic tests relevant for the axion model are presented in the Supplementary Material (SM).
}

\noindent
{\bf Axion-induced $X$-ray flux from NSs.---} The central idea behind the proposed signal is that while the cores of the M7 are quite hot ($T \sim 1 -10$ keV) the surfaces are relatively cool with $T \sim 0.1$ keV.  Axions may be emitted from hot NS interiors, escape the NSs, and then convert into hard $X$-rays in the strong magnetic fields surrounding the NSs.  To calculate the expected signal we both account for the axion production rate in the NS cores and the conversion probability in the magnetospheres.

The axions are produced in the NS cores through the axion couplings to fermionic matter.  The coupling of the axion $a$ to a fermion $\psi_f$ is denoted by (see, {\it e.g.},~\cite{Tanabashi:2018oca}) \mbox{$\mathcal{L} = (C_f / 2 f_a) \bar \psi_f \gamma^\mu \gamma_5 \psi_f \partial_\mu a$},
{with $f_a$ the axion decay constant.}
Scattering amplitudes involving this operator are generally functions of the dimensionless coupling combination $g_{aff} = C_f m_f / f_a$, with $m_f$ the fermion mass and $C_f$ the dimensionless Lagrangian coupling {(we use $C_p$ and $C_n$ for the proton and neutron, respectively).}  Note that the axion-fermion operators are generated in the infrared through the renormalization group, given an axion-photon coupling, even if such operators are absent in the ultraviolet~\cite{Srednicki:1985xd,Bauer:2017ris,Dessert:2019sgw}.

The axion production mechanisms relevant for this work mostly occur in the NS core through axion bremsstrahlung in fully degenerate nucleon-nucleon scattering $N_1 N_2 \to N_1 N_2 a$, where the $N_{1,2}$ are either neutrons or protons.  The emissivities for these processes are functions of the couplings $g_{ann}$, $g_{app}$, the local NS core temperature $T$, and the neutron and proton Fermi momenta (see the SM and~\cite{Iwamoto:1984ir,Brinkmann:1988vi}). 
 As shown in~\cite{Iwamoto:1984ir}, the local energy spectrum of axions emitted from these processes follows the modified thermal distribution $dF/dE \propto z^3(z^2 + 4 \pi^2) / (e^z - 1)$, where $z = E / T$, $E$ is the local axion energy, and $F$ is flux.
\new{We note that the nucleon bremsstrahlung rates may be suppressed at low temperatures, below the critical temperature for Cooper pair formation, by nucleon superfluidity~\cite{Yakovlev:1995kpl,Keller:2012yr}.  Recent analyses of NS cooling~\cite{Potekhin:2020ttj} indicate that the critical temperatures are likely too low to be relevant for this work, and so we ignore possible nucleon superfluidity in our fiducial analyses.  However, given that the critical temperatures are uncertain at present, we discuss their possible effects in depth in the SM.}

To compute the production rates in the NS cores, given the emissivity formulae, we need to know the temperature profiles in the cores, the metric, the critical temperature profiles \new{(if including superfluidity)}, and the profiles of neutron and proton Fermi momenta, {which all depend on the NS equation of state (EOS).}  We use the code package \texttt{NSCool}~\cite{2016ascl.soft09009P} to perform the thermal evolution of the NSs, {in full general relativity and assuming spherical symmetry.}  
 For our fiducial analysis we use the APR EOS~\cite{Akmal:1998cf} and assume NS masses of 1.4 $M_\odot$.
The thermal evolution is used to obtain a relation between the effective surface temperature and the isothermal core temperature $T_b^\infty$, which is the redshifted temperature infinitely far from the NS's potential well. \new{The surface temperatures and associated statistical uncertainties are taken from the recent compilation in~\cite{Potekhin:2020ttj}, which accounts for the effects of NS atmospheres in lowering the surface temperature for many NSs relative to the observed single-blackbody temperature.}

The relation between the surface and core temperatures is known to be strongly affected by accretion and magnetic fields, and moreover strong magnetic fields may make the surface temperature inhomogeneous (see, {\it e.g.},~\cite{Potekhin:2003aj}).  In fact it is the anisotropic surface temperatures that are thought to lead to the observed $X$-ray pulsations of the M7~\cite{PerezAzorin:2005ds}. Additionally, NS atmospheres may distort the spectra away from perfect blackbodies~\cite{Zavlin2009,Potekhin:2016gnz}. We account for these possibilities through a systematic uncertainty on the core temperatures, as described in the SM. 
We combine all $T_b^\infty$ uncertainties, statistical and systematic, into single \new{Gaussian priors, with standard deviations given in Tab.~\ref{table:NS}, with the  restriction that $T_b^\infty > 0$.}

The core temperatures may also be estimated from the kinematic ages of the NSs.  The local temperature at the outer boundary of the core $T_b$ is expected to evolve as \mbox{$T_b \approx 10^9 (t /  {\rm yr} )^{-1/6}$ K} over times $t \gg {\rm yr}$, neglecting effects such as ambipolar diffusion, which may provide additional heating to the core~\cite{Beloborodov:2016mmx}.  \new{The kinematic core temperature estimates agree with those in Tab.~\ref{table:NS} within uncertainties when the NS ages are available, though there are minor differences which, as shown in Supp.~Fig.~S8, lead to slightly lower inferred axion couplings when using core-temperature priors from age estimates. }

\begin{table}[]
\begin{tabular}{|c||c|c|c|c|c|c|}
\hline
Name  & $B_0$ & $T_s^\infty$ & $T_b^\infty$  & $d$            & $I_{x-8}$ & Refs.  \\ \hline
J0806 & $5.1$     & $100 \pm 10$    & $15 \pm 9$ & $240 \pm 25$     & $0.0_{-0.3}^{+1.6}$ & \cite{Kaplan:2009ce,Haberl:2004xe,Posselt:2006ud}   \\ \hline
J1856 & $2.9$             & $50 \pm 14$     & $5 \pm 3$  & $123 \pm 13$ & $1.5_{-0.6}^{+0.7}$ & \cite{vanKerkwijk:2007jp,Ho:2006uk,2012AA...541A..66S,Yoneyama:2017xth,2010ApJ...724..669W} \\ \hline
J0420 & $2.0$             & $45 \pm 10$     & $3 \pm 2$  & $345 \pm 200$     & $0.7_{-0.5}^{+1.0}$ & \cite{Kaplan:2011xd,Haberl:2004xe,Posselt:2006ud}   \\ \hline
J1308 & $6.8$           & $70 \pm 20$       & $8 \pm 6$  & $380 \pm 30$      & $2.3_{-1.7}^{+1.8}$ & \cite{Kaplan:2005zr,2011AA...534A..74H,2009AA...497..423M}  \\ \hline
J0720 & $6.8$           & $92 \pm 10$       & $13 \pm 8$  & $286 \pm 25$ & $0.9_{-1.6}^{+1.1}$ & \cite{Hambaryan:2017wvm,2011MNRAS.417..617T}    \\ \hline
J1605 & $2.0$             & $78 \pm 42$     & $9 \pm 11$  & $174 \pm 52$ & $-0.5_{-0.7}^{+1.3}$ & \cite{Malacaria:2019zqr,Pires:2019qsk}  \\ \hline
J2143 & $4.0$     & $72 \pm 32$     & $8 \pm 8$  & $430 \pm 200$     & $3.1_{-3.4}^{+3.0}$ & \cite{Kaplan:2009au,2009AA...499..267S,Posselt:2006ud}  \\ \hline
\end{tabular}
\caption{\label{table:NS} The M7 properties used in this work. The magnetic field strength at the pole $B_0$ is in $10^{13}$ G; the surface temperature at infinity $T_s^\infty$ is in eV; the core boundary temperature at infinity $T_b^\infty$ is in keV; the distance $d$ is in pc; the hard $X$-ray intensity $I_{x-8}$ is in $10^{-15}$ erg/cm$^2$/s, integrated from $x$ keV to 8 keV, with $x = 4$ for all NSs but J0420 and J1856, for which $x = 2$.  We obtain the NS properties from the catalog of cooling NSs~\cite{Potekhin:2020ttj}, and the $I_{x-8}$'s from Ref.~\cite{dessert2019hard}. 
}
\end{table}

{We then} consider the conversion of the axions into $X$-rays in the NS magnetic fields.  Here we follow closely the framework outlined in~\cite{Dessert:2019sgw} for axion-photon conversion in WD magnetospheres.  The axion-photon mixing is induced through the operator \mbox{$\mathcal{L} =- g_{a\gamma\gamma} a F \tilde F/4$}, where $F$ is the electromagnetic field strength tensor, $\tilde F$ is its dual field, and $g_{a\gamma\gamma}$ is the axion-photon mixing parameter.  The parameter $g_{a\gamma\gamma}$ is related to $f_a$ through the relation $g_{a\gamma\gamma} = C_\gamma \alpha_{\rm EM} / (2 \pi f_a)$, with $ \alpha_{\rm EM}$ the fine structure constant and $C_\gamma$ a dimensionless coupling constant. 
  In the presence of a strong magnetic field this operator may cause an initially pure axion state to rotate into an electromagnetic wave polarized parallel to the external magnetic field.  However, the axion-photon conversion is suppressed by the Euler-Heisenberg term for strong field quantum electrodynamics~\cite{Raffelt:1987im}.  
 
{In the limit of low axion mass, which for our applications is roughly $m_a \lesssim (\omega R_{\rm NS}^{-1})^{1/2}$ (and approximately $10^{-4}$ eV at axion frequencies $\omega \sim {\rm keV}$ and NS radii $R_{\rm NS} \sim 10$ km), the conversion probability $p_{a \to \gamma}$ is approximately~\cite{Raffelt:1987im,Fortin:2018ehg,Dessert:2019sgw} }
\es{conv_prob_low_mass}{
p_{a \to \gamma} \approx & 1.5 \times 10^{-4} \left( {g_{a \gamma \gamma} \over 10^{-11} \, \, {\rm GeV}^{-1}} \right)^2 \left( {1 \, \, {\rm keV} \over \omega} \right)^{4/5} \\
 &\left( {B_0 \over 10^{13} \, \, {\rm G}} \right)^{2/5} \left( {R_{\rm NS} \over 10 \, \, {\rm km} }\right)^{6/5} \sin^{2/5}\theta \,,
}
independent of the axion mass.  Above, $B_0$ is the surface magnetic field strength at the magnetic pole, {assuming a dipole field configuration}, and $\theta$ is the polar angle from the magnetic axis. 
  At large axion masses the conversion probability becomes additionally suppressed and must be computed numerically (see, {\it e.g.},~\cite{Dessert:2019sgw}). 

We assume dipolar magnetic field strengths calculated from the spindown of the NSs~\cite{Kaplan:2009ce,vanKerkwijk:2007jp,Kaplan:2011xd,Kaplan:2005zr,Kaplan:2005fy,Hambaryan:2017wvm,Kaplan:2009au} via magneto-dipole radiation. \new{(Note that the statistical uncertainties on the dipole field strengths are sub-leading.)} In the case of J1605, there is no spin-down measurement and we adopt $2 \times 10^{13}$ G as considered in~\cite{Malacaria:2019zqr}. Measurements of the magnetic field from spectral fitting of cyclotron resonance lines or atmosphere models generally predict larger fields, which we consider in the SM.  
\new{We account for the unknown alignment angle $\theta$ by profiling over $\theta$ with a flat prior. }

\noindent
{\bf Data analysis.---} \new{Ref.~\cite{dessert2019hard} analyzed all available archival data from {\it XMM-Newton} and {\it Chandra} towards each of the M7 for evidence of hard $X$-ray emission.  For {\it XMM-Newton} they reprocessed data from both the MOS and PN cameras and treated these datasets independently, since they are subject to different sources of uncertainty from {\it e.g.}~pileup.  The data were binned into three high-energy bins from 2 - 4, 4 - 6, and 6 - 8 keV.  Ref.~\cite{dessert2019hard} computed likelihood profiles for flux from the M7 in each one of these energy bins; these likelihoods are provided as Supplementary Data in~\cite{dessert2019hard} and are the starting points for the analyses presented in this work.}  As an illustration, in Fig.~\ref{fig:example_spec}
\begin{figure}[htb]
\hspace{0pt}
\vspace{-0.3in}
\begin{center}
\includegraphics[width=\columnwidth]{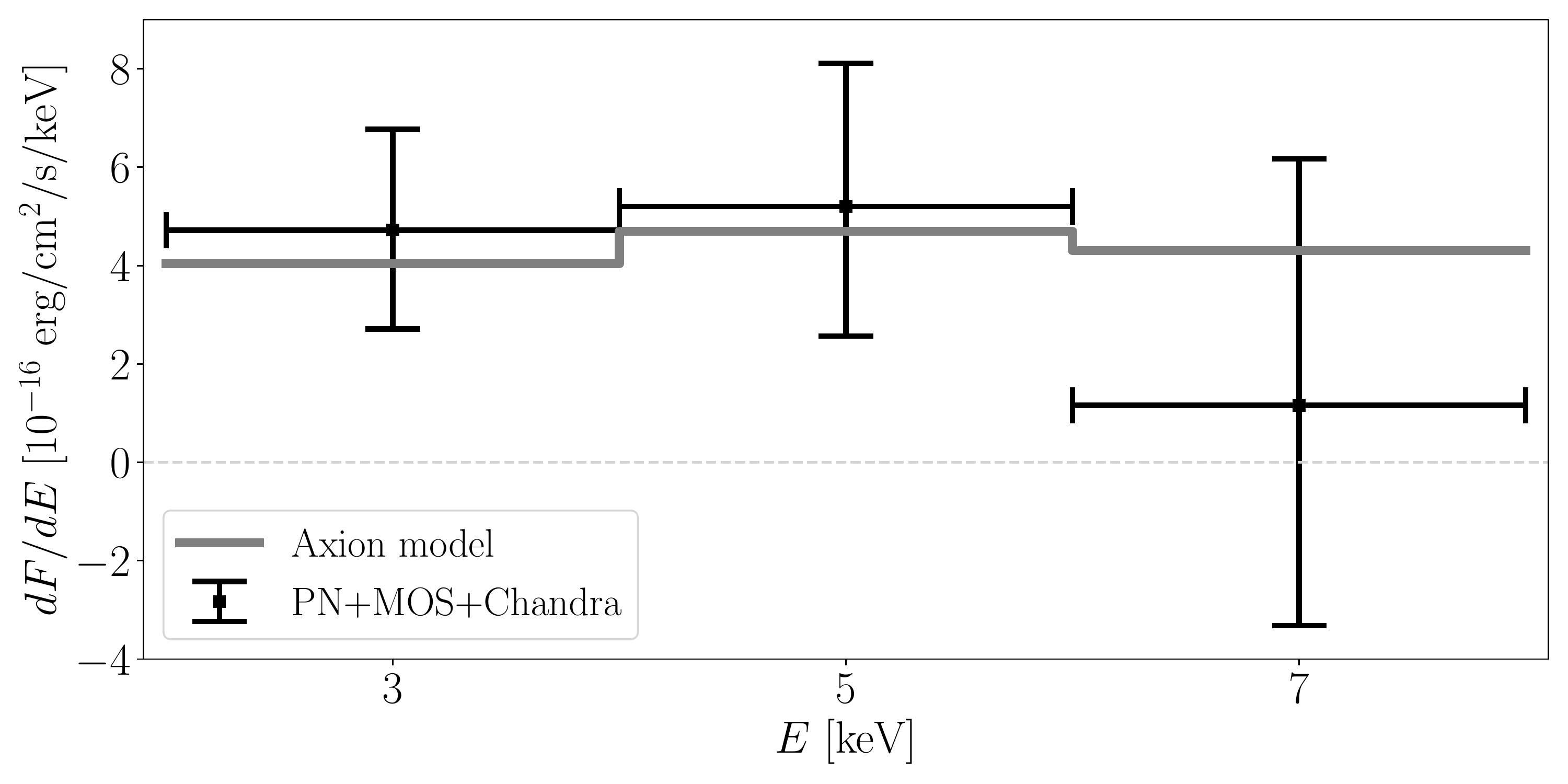}
\caption{
\label{fig:example_spec}
The energy spectrum from 2 to 8 keV for NS J1856 as measured by combining PN, MOS, and {\it Chandra} data, with 68\% statistical uncertainties~\cite{dessert2019hard}. 
We also show the best-fit axion model spectrum from a fit to this NS only, with the core temperature fixed to the central value in Tab.~\ref{table:NS}.}
\end{center}
\end{figure}
 we show the energy spectrum from J1856, which is the NS with the most significant hard $X$-ray excess. Note that we show the best-fit fluxes and associated 68\% confidence intervals from the joint analyses over all three cameras.
 
 \new{
Ref.~\cite{dessert2019hard} showed that the 2 -- 4 keV energy bin may be contaminated by the high-energy tail of the thermal emission from the NS surfaces, depending on the atmosphere model, for all NSs except J1856 and J0420.  The predicted thermal surface emission is negligible for all NSs in the last two energy bins.  As such in this analysis we use all three available energy bins for J1856, which has by far the most exposure time of all M7, and J0420, but only the last two energy bins for the other five NSs. 
Ref.~\cite{dessert2019hard} only provides {\it Chandra} data for  J1856, J0420, and J0806, because for the other NSs they found that pileup may affect the observed high-energy spectrum.  For J2143 only PN data is available.
}  

We interpret the M7 hard $X$-ray spectra in the context of the axion model by using a joint likelihood over all of the M7 and available datasets \new{with a frequentist profile likelihood analysis} procedure.  Our parameters of interest are $\{m_a, g_{a\gamma\gamma}, g_{ann}, g_{app}\}$ and our nuisance parameters, which describe uncertain aspects of the NSs, are the set of parameters $\{ \theta, d, T_b^\infty\}$ for each NS, where $d$ is distance.
  Each of the nuisance parameters is taken to have a Gaussian prior with uncertainty given in Tab.~\ref{table:NS}, except for $\theta$, which is given a flat prior from $0$ to $\pi$. Uncertainties arising from the NS superfluidity model are described in the SM.  For our fiducial analysis we fix $g_{app} = g_{ann}$. 
  We construct a joint likelihood over all of the M7 and available datasets, and we use this likelihood to constrain our parameters of interest. 

  \noindent
  {\bf Results.---}
The resulting best-fit parameter space in the $m_a$-$g_{a\gamma\gamma} g_{ann}$ plane and 95\% one-sided upper limit are shown in Fig.~\ref{fig:money}.  
\begin{figure}[htb]
\hspace{0pt}
\vspace{-0.2in}
\begin{center}
\includegraphics[width=0.5\textwidth]{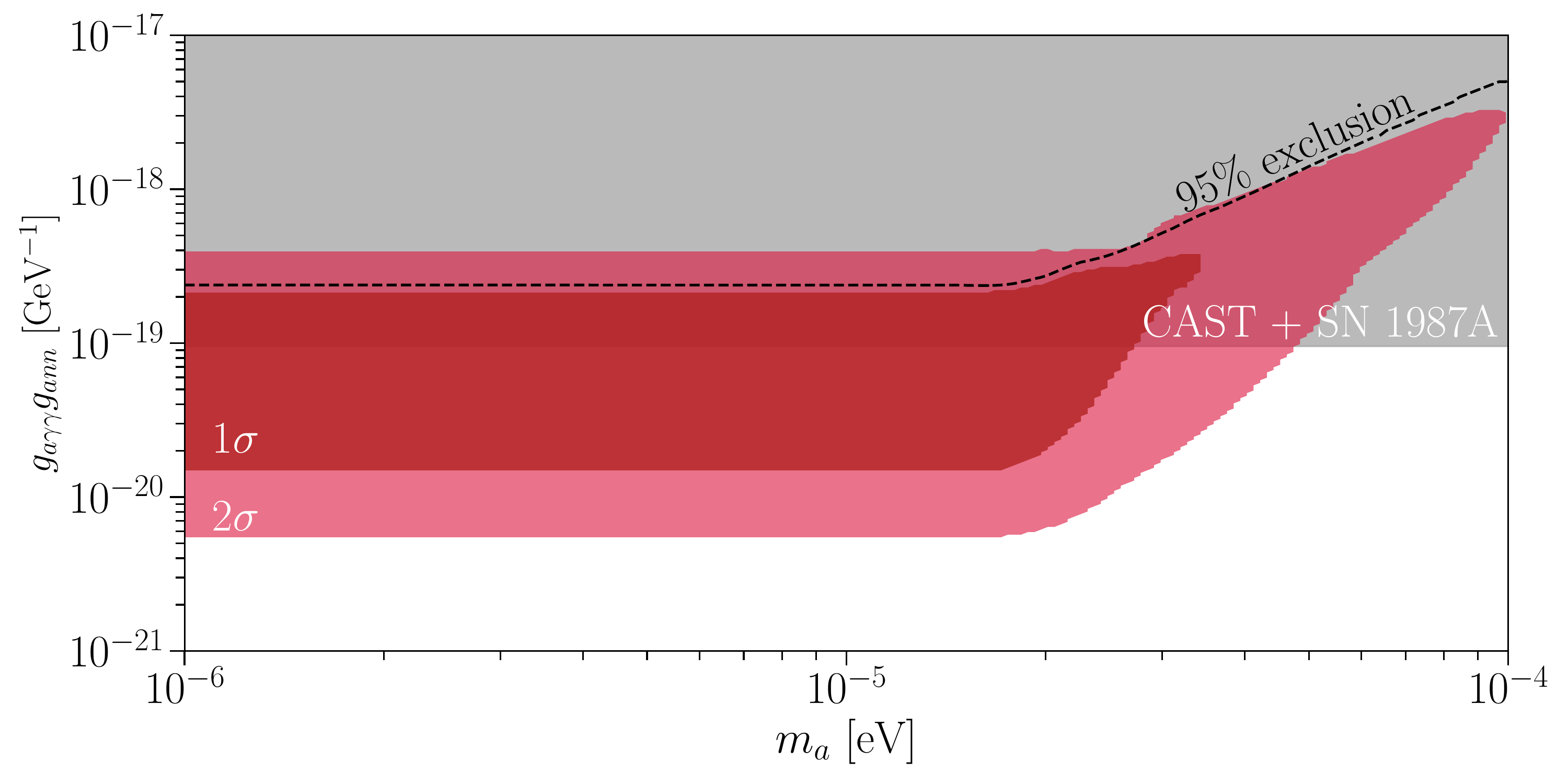}
\caption{
\label{fig:money}
95\% exclusion limit and best fit 1 and 2$\sigma$ regions from a joint likelihood analysis over all of the M7 and combining PN, MOS, and $\it Chandra$ data. We compare our result to existing limits from CAST2017+NS cooling.
  All curves and regions continue to arbitrarily small $m_a$.  Note that the QCD axion model is too weakly coupled to appear in this figure.  \new{Accounting for systematic uncertainties may allow for smaller values of $g_{a\gamma\gamma} g_{ann}$, by approximately an order of magnitude, as discussed in the SM.} 
}
\end{center}
\end{figure}
Interpreting the data in the context of the axion model, we find approximately $5$$\sigma$ evidence for the axion-induced flux over the null hypothesis of no non-thermal hard $X$-ray flux from the M7. 
The global fit prefers a low axion mass and a coupling at and slightly below previous limits, which are also indicated. In particular we combine the CAST constraints on $g_{a\gamma\gamma}$ ($g_{a\gamma\gamma} < 6.6\times10^{-11}$ GeV$^{-1}$ at low masses)~\cite{Anastassopoulos:2017ftl} with \new{the SN 1987A constraints on $g_{ann}$, taking $g_{app} = g_{ann}$, $g_{ann} < 1.4\times10^{-9}$~\cite{Zyla:2020zbs} (but see~\cite{Bar:2019ifz} which questions these constraints).  Constraints on $g_{ann}$ from cooling of the NS Cas A~\cite{Keller:2012yr,Sedrakian:2015krq,Sedrakian:2018kdm} may all be relevant, though these constraints are subject to both instrumental~\cite{Posselt:2018xaf} and theoretical systematic uncertainties.  Thus the current constraints on $g_{a\gamma\gamma} g_{ann}$ in Fig.~\ref{fig:money} should be taken as suggestive. }

It is interesting to investigate whether the high-energy flux observed between the individual NSs is consistent with the expectation from the axion hypothesis.  In Fig.~\ref{fig:I28} we show the observed intensities $I_{2-8}$ ($I_{4-8}$) between 2 - 8 keV (4 - 8 keV) for each of the M7 after combining the MOS, PN, and {\it Chandra} datasets.  These intensities are determined by fitting the low-mass axion spectral model uniquely to the data from each NS, with model parameters $T_b^\infty$ and $I_{2-8}$ ($I_{4-8}$).  Note that for the NSs where we include the 2 - 4 keV energy bin we report $I_{2-8}$, while for those where we do not include this bin we instead report $I_{4-8}$. (We obtain qualitatively similar results if we only use the 4 - 8 keV bins for all NSs, as shown in the SM.)   The green (yellow) bands indicate the 68\% (95\%) confidence intervals for the intensities from the $X$-ray measurements.  The black and gray error bands, on the other hand, denote the 68\% and 95\% confidence intervals for the axion model predictions, fixing the axion model parameters at the best-fit point from the global fit, $g_{a\gamma\gamma} g_{ann} \approx 4 \times 10^{-20}$ GeV$^{-1}$ with $m_a \ll 10^{-5}$ eV, and profiling over the nuisance parameters. 
The uncertainties in the model prediction arise primarily from the nuisance parameters describing the unknown properties of the M7, as described above, while the uncertainties on the measured intensity values are purely statistical in nature.  

The observed intensities are consistent with expectations from the axion model.
Additionally, there are sources of uncertainty on the axion model predictions for the individual NSs beyond those shown  in Fig.~\ref{fig:I28}, arising from for example nucleon superfluidity, the EOS, and the inference of the core temperatures.  \new{For example, as we show in the SM with alternate core-temperature models, based on ages rather than surface temperatures, the best-fit couplings could be as low as $g_{a\gamma\gamma} g_{ann} \approx 2 \times 10^{-21}$ GeV$^{-1}$.}
\begin{figure}[htb]
\hspace{0pt}
\vspace{-0.2in}
\begin{center}
\includegraphics[width=0.5\textwidth]{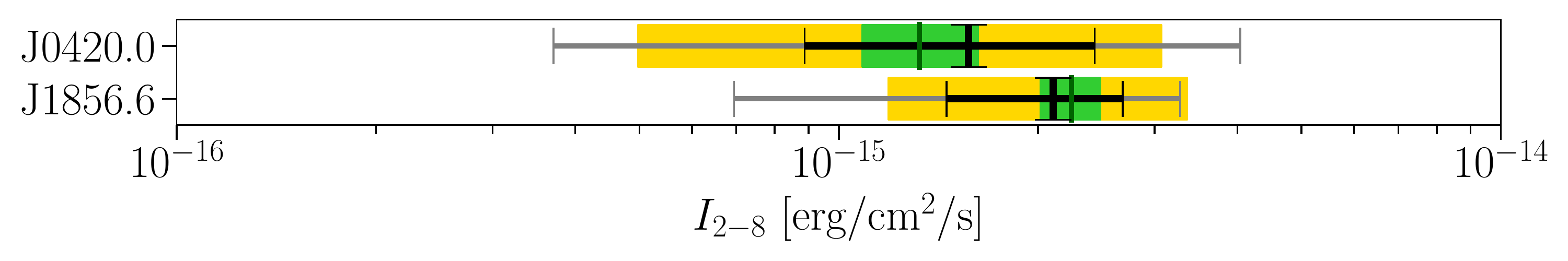}
\includegraphics[width=0.5\textwidth]{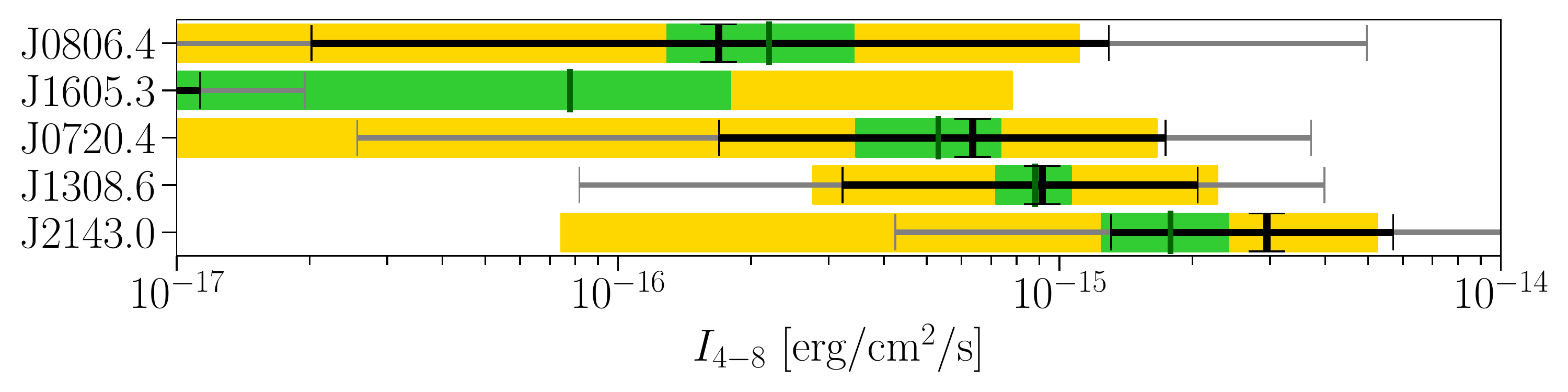}
\caption{
\label{fig:I28}
Best-fit intensities $I_{2-8}$ and $I_{4-8}$ for all M7. 
The green (yellow) bands indicate the 68\% (95\%) confidence intervals from the $X$-ray intensity measurements, with best-fit intensities marked by vertical green lines.  Black and gray error bands denote the 68\% and 95\% confidence intervals for the axion model predictions at the global best-fit coupling $g_{a\gamma\gamma} g_{ann}$ and $m_a \ll 10^{-5}$ eV, with uncertainties arising from uncertain aspects of the NSs.
}
\end{center}
\end{figure}

We also investigate whether the observed spectra from the two high-significance detections in J1856 and J0420 are consistent with the axion model expectation. In Fig.~\ref{fig:Ts} we show the best-fit core temperatures $T_b^\infty$ measured from fitting the axion-model, with $m_a \ll 10^{-5}$ eV, to the $X$-ray data between 2 and 8 keV. 
\begin{figure}[htb]
\hspace{0pt}
\vspace{-0in}
\begin{center}
\includegraphics[width=0.5\textwidth]{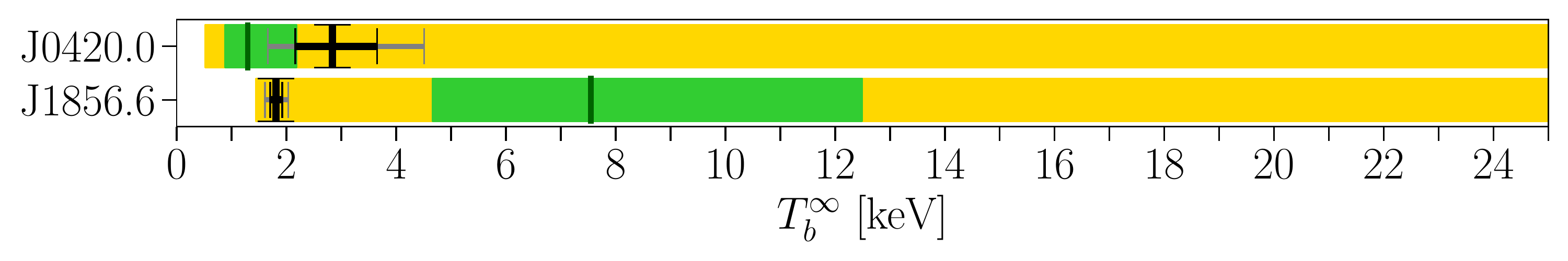}
\caption{
\label{fig:Ts}
{As in Fig.~\ref{fig:I28} but for the best-fit core temperature $T_b^\infty$ for J1856 and J0420.} 
}
\end{center}
\end{figure}
We note that the NS with the best-determined spectral shape is J1856, which has the most significant detection.  In Fig.~\ref{fig:example_spec} we show the best-fit model prediction for this NS compared to the observed spectrum.  The axion model appears to reproduce the spectral shape found in the data.  

\noindent
{\bf Discussion.---}
In this work we presented results of a search for hard $X$-ray emission arising from axions in the M7 NSs.  
\new{We showed that the M7 hard $X$-ray excess observed in~\cite{dessert2019hard} may be interpreted in the context of the axion model. } 

Alternative explanations for the hard $X$-ray emission exist, but they are not compelling~\cite{dessert2019hard}.  For example, some of the observations may be affected by pileup due to the high flux of soft, thermal $X$-rays, though these effects seem insufficient to explain the observed hard $X$-ray flux~\cite{dessert2019hard}.  For the {\it XMM-Newton} data in particular, unresolved astrophysical point sources near the source of interest could also bias the observed spectrum, though the fact that \new{consistent spectra are observed with} {\it Chandra}, which has over an order of magnitude better angular resolution, provides evidence that this is at least not the sole explanation for the excess.  Hard non-thermal $X$-ray emission is observed generically from pulsars, and one possibility is that the observed hard $X$-ray flux from the M7 arises from the traditional non-thermal emission mechanisms ({\it e.g.}, synchrotron emission) that are present in other pulsars.  On the other hand, this emission is often accompanied by non-thermal radio emission, which is not observed for the M7~\cite{2009ApJ...702..692K}, and also the spin-down luminosity seems insufficient for most of the M7 for this to be an appreciable source of flux~\cite{dessert2019hard}.  Accretion of the interstellar medium may also be a source of $X$-rays from the M7, though this is typically thought to produce flux at much softer energies if at all (see, {\it e.g.},~\cite{Treves:1999ne}). 

Observations at higher energies by {\it e.g.~NuSTAR} of J1856 and J0420 in particular may help discriminate the axion explanation of the excess from other explanations.  This is because the predicted axion spectrum in the energy range from $\sim$10-60 keV is unique and potentially includes a significant enhancement due to Cooper pair-breaking-formation processes, depending on the superfluidity model. (See the SM for details, where we also show that the Cooper pair processes could enhance the flux below 10 keV as well.) 
The axion-induced flux should also pulsate at the NS period, and this may be measurable with future instruments such as {\it Athena} that can acquire better statistics. {\it Athena} will have similar angular resolution to {\it Chandra} while also being significantly less affected by pileup~\cite{Barret:2016ett}.  
{$X$-ray observations of magnetic white dwarf stars~\cite{Dessert:2019sgw}, the magnetized intracluster medium~\cite{Reynolds:2019uqt}, or nearby bright galaxies~\cite{Marsh:2017yvc} could also help constrain or provide additional evidence for the best-fit axion from this work.}
{The best-fit axion parameter space from this work may also be probed with next-generation light-shining-through-walls experiments like ALPS II~\cite{Spector:2016vwo} and helioscopes like IAXO~\cite{Armengaud:2019uso}.  }
  In summary, if the M7 hard $X$-ray excess is due to axions, then a variety of near-term measurements should be able to conclusively establish a discovery.

\begin{acknowledgments}
{\it 
We are grateful to Y. Kahn for collaboration in the early stages of this work and comments on the manuscript and to J. Foster, M. Reynolds, O. Gnedin, H. G{\"u}nther, D. Hooper, A. Long, A. Ringwald, and D. Yakovlev for useful discussions and comments. We thank the anonymous referees for useful suggestions. This work was supported in part by the DOE Early Career Grant DE-SC0019225 and the DOE grant DE-SC0011842 (R.C.) at the University of Minnesota and through computational resources and services provided by Advanced Research Computing at the University of Michigan, Ann Arbor. CD was partially supported by the Leinweber Graduate Fellowship at the University of Michigan, Ann Arbor. This work was performed in part at the Aspen Center for Physics, which is supported by the National Science Foundation grant PHY-1607611, and in part at the Mainz Institute for Theoretical Physics (MITP) of the Cluster of Excellence PRISMA+ (Project ID 39083149).  We also acknowledge the Munich Institute for Astro- and Particle Physics (MIAPP) of the DFG Excellence Cluster Origins along with the CERN Theory department for hospitality during the completion of this work. 
}
\end{acknowledgments}

\bibliography{axion}

\clearpage
\newpage
\maketitle
\onecolumngrid
\begin{center}
\textbf{\large Axion Emission can Explain a New Hard $X$-ray Excess from Nearby Isolated Neutron Stars} \\ 
\vspace{0.05in}
{ \it \large Supplementary Material}\\ 
\vspace{0.05in}
{Malte Buschmann, \ Raymond T. Co, \ Christopher Dessert, \ and \ Benjamin R. Safdi}
\end{center}
\onecolumngrid
\setcounter{equation}{0}
\setcounter{figure}{0}
\setcounter{table}{0}
\setcounter{section}{0}
\setcounter{page}{1}
\makeatletter
\renewcommand{\theequation}{S\arabic{equation}}
\renewcommand{\thefigure}{S\arabic{figure}}
\renewcommand{\thetable}{S\arabic{table}}

This Supplementary Material is organized as follows. Section~\ref{sec:NS_Tb} discusses our determination of the core temperature uncertainties, given the surface temperature data for the M7. In Sec.~\ref{SM:brem_rates} we outline our computation of the axion production rates via nucleon bremsstrahlung accounting for the possible suppression of the rates during neutron superfluidity.  \new{Sec.~\ref{sec:stats} details the statistical analysis framework used to interpret the $X$-ray data in the context of the axion model.} In Sec.~\ref{SM:PBF} we present calculations of the axion emission rate and spectrum via the Cooper pair-breaking-formation (PBF) processes and discuss the expected spectra from the NSs. Finally in Sec.~\ref{SM:sysTest} we perform multiple systematic tests on the analyses presented in the main text and discuss the robustness of our results.

\section{Core Temperatures}
\label{sec:NS_Tb}

In this section, we estimate the uncertainties in the determinations of the core temperatures from the known surface temperatures of the NSs. The inner region of the NS is isothermal in the sense that the redshifted temperature observed infinitely far from the NS, $T_b^{\infty} = T_b (r) \sqrt{g_{00}(r)}$, is independent of the production radius $r$ within the NS, with $g_{00}$ the temporal component of the metric.    We define the un-redshifted core temperature as $T_b = T_b(r_b)$, where $r_b$ is the radius of the outer boundary of the isothermal internal region.  Note that $r_b$ is slightly smaller than the radius of the NS, $r_{\rm NS}$. Surrounding the isothermal region is the NS envelope, over which the temperature cools to the surface temperature $T_s = T_s^\infty / \sqrt{g_{00}(r_{\rm NS})}$ at the outer surface. 

In practice, we use \texttt{NSCool} to compute $T_b$ given $T_s$, but we estimate the uncertainty in this determination using the analytic relations determined from fits to simulations given in Eq.~32 of~\cite{1983ApJ...272..286G} and Sec.~A.3 of~\cite{Potekhin:1997mn}. The majority of the uncertainty arises from the uncertainty in the NS surface gravity (because of the uncertainty in the NS EOS) and in the NS accretion history.  The NS EOS and the NS masses are sources of uncertainty that should be more thoroughly investigated in future work.

\begin{figure}[htb]
\includegraphics[width=0.5\linewidth]{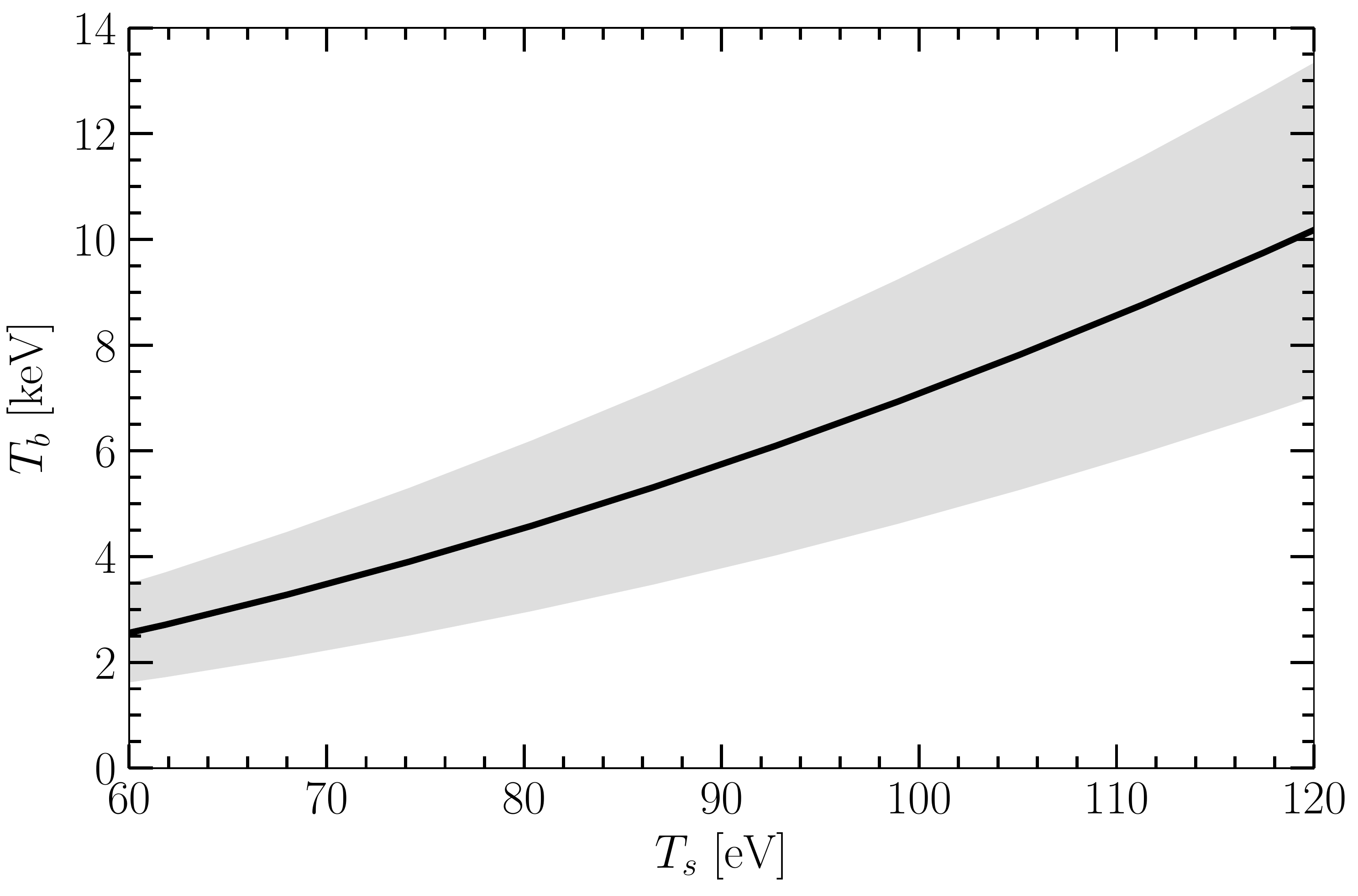}
\caption{An illustration of the uncertainty on the determination of $T_b$, given the surface temperature $T_s$, arising from the uncertainties in the surface gravity and the amount of accreted matter. The black curve shows the average value of $T_b$ for each given $T_s$ if one assumes flat priors in $2 \le g_{14} \le 6$ and $-20 \le \log(M_{\rm ac}/M_{\rm tot}) \le -10$, whereas the gray band shows the 68\% containment region on $T_b$ given $T_s$.}
\label{fig:Tb_Ts}
\end{figure}

We estimate such uncertainties by varying over the amount of accreted matter $M_{\rm ac}$ and the surface gravity $g_{14}$ in $10^{14}$ cm/s$^2$. We assume a wide range of $2 \le g_{14} \le 6$, as a conservative estimate based on~\cite{Bejger:2004gz}. The NSs of interest are isolated and are not expected to accrete much matter. We assume a flat prior in $-20 \le \log(M_{\rm ac}/M_{\rm tot}) \le -10$, where $M_{\rm tot}$ is the total mass of the NS. In the relevant range of surface temperatures, we find that the standard deviation of $T_b$ is around 30\% of the mean.  This is illustrated in Fig.~\ref{fig:Tb_Ts}, where we show the central $T_s$-$T_b$ relation along with the 68\% containment region from varying over $M_{\rm ac}$ and $g_{14}$ as described above. \new{In the analyses throughout this work, we use normal uncertainties on the core temperatures, accounting for the 30\% systematic uncertainty.}  

Additional uncertainty on the core temperature arises from the intrinsic uncertainty in the surface temperatures. In addition to the uncertainty determined in the Ref.~\cite{dessert2019hard}, we assign a $25\%$ uncertainty on the surface temperature to account for the variation in NS atmosphere models as well as the unknown surface composition. Note that the surface composition is related to the accretion history and so some of the uncertainties are interrelated. We convert the surface temperature uncertainties to core temperature uncertainties using the analytic relations above. We finally combine the three uncertainties into a single normal prior on the core temperature reported in Tabs.~\ref{table:NS} and~\ref{tab:altNS}.  Lastly, we note that as mentioned in the main text the core temperatures may also be estimated from the kinematic ages of the NSs, and when such estimates are available we find good agreement, within uncertainties, between the surface-temperature-based core temperature estimates and the age-based estimates of the core temperatures.  

\section{Nucleon bremsstrahlung rates}
\label{SM:brem_rates}

Here we provide a brief overview of how the nucleon bremsstrahlung rates are calculated in the NS cores.
We can safely assume a degenerate limit for the nucleon-nucleon bremsstrahlung emission in the NS cores because the NSs we consider have $T \ll \mathcal{O}$(MeV)~\cite{Brinkmann:1988vi}. The production rate of axions from a NS via nucleon-nucleon bremsstrahlung emission in the degenerate limit is calculated by~\cite{Iwamoto:1984ir, Iwamoto:1992jp}, assuming no nucleon superfluidity.

These production modes are suppressed by an energy-independent factor below the critical temperature $T_c$ at which the nucleons form Cooper pairs and undergo a phase transition into the superfluid state.  Note that Cas A cooling measurements provide indirect evidence that such a transition takes place~\cite{Page:2010aw}.
 This is because the superfluid suppression also occurs for neutrino emission via the modified Urca process, and Cas A is thought to cool primarily from neutrino emission~\cite{Yakovlev:1995kpl}.  The explicit formulae that we use for the  singlet-state pairing suppression factors  may be found in~\cite{Yakovlev:1995kpl} (see also~\cite{Keller:2012yr}). 
 The neutrons in the core, however, are thought to undergo triplet-state pairing, and the explicit formula for the triplet-state pairing suppression factors have not been worked out.  We follow the code package \texttt{NSCool} and approximate the triplet-state pairing suppression factors by the singlet-state ones~\cite{2016ascl.soft09009P}.

\section{Statistical analysis framework}

\label{sec:stats}

In this section we briefly overview the statistical analysis framework that we use to interpret the $X$-ray data in the context of the axion model. 
\new{Our starting point is the photon-count based likelihoods computed in~\cite{dessert2019hard}.  These are given by functions $\mathcal{L}_{i}({\bm x}_i|S_i)$ of flux $S_i$, where $i$ labels the energy bin and ${\bm x}_i$ denotes the dataset associated with that energy bin.  These Poisson likelihoods use the expected number of background counts in the signal region, the observed number of counts, and the conversion factor to go from flux to counts accounting for the instrumental response; all of these quantities are provided in~\cite{dessert2019hard}.}

Given the likelihoods in the individual energy bins we may construct the joint likelihood that constrains the axion model:
\es{}{
{\mathcal L}_{\rm axion}({\bm x}| \{ m_a, g_{a\gamma\gamma} g_{ann}, {\bf \theta_n} \}) = \prod_i  \mathcal{L}_{i}\big[ {\bm x}_i| S_i(\{ m_a, g_{a\gamma\gamma} g_{ann}, {\bf \theta_n} \})\big] \times \mathcal L_{\rm prior}({\bf \theta_n}) \,,
} 
where the product $i$ is over all energy bins used in the analysis and where $\mathcal{L}_{\rm prior}({\bf \theta_n}) $ denotes the prior distribution for the nuisance parameters $\bf \theta_n$, which will be discussed more shortly.  The axion model parameters are the mass $m_a$ and the coupling combination $g_{a\gamma\gamma} g_{ann}$, while the dataset ${\bm x} = \{ {\bm x}_i \}$ is the union of the datasets in the individual energy bins.  The nuisance parameters ${\bf \theta_n} = \{d, T_b^\infty, \theta\}$ denote the uncertain properties of the NSs that we vary over in the fit, such as the distance $d$, the core temperature $T_b^\infty$, and the alignment angle with respect to Earth $\theta$.  In particular we use the (un-normalized) prior distribution function 
\es{}{
\mathcal{L}_{\rm prior}({\bf \theta_n}) =   \exp \left[ - (T_b^\infty - \bar T_b^\infty)^2 \over 2 \sigma_{ T_b^\infty}^2 \right]  \exp \left[ - (d - \bar d)^2 \over 2 \sigma_{d}^2 \right] \Theta(T_b^\infty) \Theta(d) \Theta(\theta \times (\pi-\theta)) \,,
}
where quantities with an over-bar denote the central measured parameters given in Tab.~\ref{table:NS}, the $\sigma$'s denote the standard deviations presented in that table, and $\Theta$ is the Heaviside step function so that $T_b^\infty$ and $d$ stay positive and $\theta$ has a flat prior between $(0, \pi)$.  

To construct 95\% upper limits on $g_{a\gamma\gamma} g_{ann}$ we fix $m_a$ and consider the profile likelihood ${\mathcal L}_{\rm axion}({\bm x}| \{ m_a, g_{a\gamma\gamma} g_{ann} \})$ as a function of $g_{a\gamma\gamma} g_{ann}$.  Note that to construct the profile likelihood we maximize the log-likelihood over the nuisance parameters.  We employ Wilks' theorem to assume that the log-likelihood is asymptotically $\chi^2$ distributed (we have checked that this is valid explicitly with Monte Carlo), so that the 95\% upper limit may be found from the value $g_{a\gamma\gamma} g_{ann} > \overline{g_{a\gamma\gamma} g_{ann}}$ such that $2 \times \left[ {\mathcal L}_{\rm axion}({\bm x}| \{ m_a, g_{a\gamma\gamma} g_{ann}\}) - {\mathcal L}_{\rm axion}({\bm x}| \{ m_a, \overline{ g_{a\gamma\gamma} g_{ann} } \})  \right] \approx -2.71$ (see {\it e.g.}~\cite{Cowan:2010js}), with $\overline{ g_{a\gamma\gamma} g_{ann} }$ denoting the coupling combination that maximizes the likelihood at fixed $m_a$.  To search for evidence of the axion mode, which we find, we consider the discovery test statistic ${\rm TS} =  - 2 \times \left[ {\mathcal L}_{\rm axion}({\bm x}| \{ m_a, g_{a\gamma\gamma} g_{ann}\}) - {\mathcal L}_{\rm axion}({\bm x}| \{ m_a, \overline{ g_{a\gamma\gamma} g_{ann} } \})  \right]$, which is a function of both $m_a$ and $g_{a\gamma\gamma} g_{ann}$.  The best-fit $1\sigma$ interval, for example, is defined by the region in axion mass and coupling space where the discovery TS is within a value of unity from the maximum, again assuming Wilks' theorem holds (which we have checked explicitly).

\section{Cooper Pair-Breaking-Formation Processes}
\label{SM:PBF}

Cooper pairs are expected to form when the temperature is below the superfluid critical temperature.
When the temperature is still not far below the critical temperature, the thermal interactions can break the Cooper pairs. Neutrinos and axions can be produced and carry away energy released during these Cooper pair breaking and formation processes. \new{In the fiducial analysis, this production mode was not active because our fiducial analysis assumes no core superfluidity.}
In this section we review the axion emission rates from these processes, derive the energy spectrum, and discuss the implication for the high-energy $X$-ray flux.

\subsection{Emission Rates}
\label{SM:PBF_rates}

The NS cores may contain spin-$0$ $S$-wave and spin-$1$ $P$-wave nucleon superfluids.  
There then exists a production mode of axions via Cooper pair-breaking-formation (PBF), with a rate for the $S$-wave pairing given by~\cite{Keller:2012yr, Sedrakian:2015krq}
\es{eq:PBF_S}{
\epsilon_{a,{\rm PBF}}^S  &= \frac{2 g_{aNN}^2}{3\pi m_N^2} \,\nu_N(0)\, v_F(N)^2 \, T^5 \, I_a^S \,, \\ 
 I_a^S &= z_N^5\int_1^{\infty}\!\! dy ~ \frac{y^3}{\sqrt{y^2-1}} \left[ f_F\left(z_N   y\right)\right]^2  \,,  \qquad {\rm with} \ \ f_F(x)  = \frac{1}{e^x+1} \,.}
Above, $\nu_N(0) = m_N p_F(N)/\pi^2$ is the density of states at the Fermi surface, with $v_F(N)$ the fermion velocity. Here $z_N= \Delta(T)/T$, and a simple analytic fit for the superfluid energy gap $\Delta(T)$ is given in \cite{Yakovlev:1995kpl}. 
The PBF process is active when the temperature falls below the critical temperature $T_c$. Due to the sensitive dependence of $T$ in $f_F$, the emission rate is exponentially suppressed however at low temperatures, {\it i.e.}~$T \ll T_c$. 

One should identify $2 y \Delta(T)$ as the axion energy $\omega$ (this follows from the derivation of~\eqref{eq:PBF_S}), and thus the energy spectrum follows the functional form
\begin{equation}
\label{eq:JaS}
J_{a,{\rm PBF}}^S \equiv \frac{d\left( \epsilon_{a,{\rm PBF}}^S \right)}{d \omega} = \frac{\mathcal{N}_{a,{\rm PBF}}^S}{2\Delta(T)}   \frac{ \left(\frac{\omega}{2\Delta (T)} \right)^3}{\sqrt{\left(\frac{\omega}{2\Delta (T)} \right)^2-1}} \left[ f_F \left( \frac{\omega}{2T}\right) \right]^2 ,
\end{equation}
where $\mathcal{N}_{a,{\rm PBF}}^S$ is the normalization constant determined by $\int_{2\Delta (T)}^\infty J_{a,{\rm PBF}}^S d\omega = \epsilon_{a,{\rm PBF}}^S$ and reads $\mathcal{N}_{a,{\rm PBF}}^S = \epsilon_{a,{\rm PBF}}^S z_N^5/I_a^S$.
Here $T$ and $\omega$ refer to the locally-measured quantities inside the NS at some radius $r_0$, {\it i.e.}~$T = T_b(r_0) = T_b^\infty / \sqrt{g_{00}(r_0)}$ and $\omega = \omega(r_0) = \omega_\infty / \sqrt{g_{00}(r_0)}$. Practically, one first computes the initial spectral function $J_{a,{\rm PBF}}^S (\omega(r_0))$ at each radius $r_0$ using the local temperature $T_b(r_0)$ and then interprets the observed spectral function as $J_{a,{\rm PBF}}^S ( \omega_\infty / \sqrt{g_{00}(r_0)})$ with $\omega_\infty$ identified as the observed $X$-ray energy.

Similarly, the rate for the neutron $P$-wave pairing is~\cite{Keller:2012yr, Sedrakian:2015krq}
\es{eq:IanP}{
\epsilon_{a,{\rm PBF}}^P  &= \frac{2 g_{ann}^2}{3\pi m_N^2} \,\nu_n(0) \, T^5 \, I_{an}^P \,, \\
I^P_{an} (z_x) &= \int \frac{d\Omega}{4\pi} z_x^5\int_1^{\infty}\!\! dy ~ \frac{y^3}{\sqrt{y^2-1}} \left[f_F\left(z_x y\right)\right]^2 \,, \qquad {\rm with} \ \ f_F(x)  = \frac{1}{e^x+1} \,.
}
There exist two types of the $P$-wave pairings. In \cite{Sedrakian:2015krq}, types $A$ and $B$ refer to the $^3P_2$ pairing with total projection of the Cooper-pair momentum onto the z-axis equal to $m_J = 0$ and $2$, respectively. The anisotropic superfluid gaps are given by 
\es{eq:DeltaT_PA}{
\Delta_A (T, \theta) & = \Delta_0^{(A)}(T) \sqrt{1+3 \cos^2\theta} \,, \\
\Delta_B (T, \theta) & = \Delta_0^{(B)}(T) \sin\theta \,,
}
with $\theta$ the angle between the neutron momentum and the quantization axis and $z_x \equiv \Delta_x (T, \theta)/T$.  Explicit expressions for $\Delta_0^{(A,B)}(T)$ may be found in~\cite{Sedrakian:2015krq} along with approximations for the phase space integrals $I_{an}^P$.

With $2 y \Delta_x(T, \theta)$ identified as the axion energy $\omega$, the spectra for the $P$-wave pairings follows as
\begin{align}
\label{eq:JaP}
J_{a,{\rm PBF}}^P & \equiv \frac{d\left( \epsilon_{a,{\rm PBF}}^P \right)}{d \omega} = \int_{-1}^{1} \frac{d\cos\theta}{2} \frac{1}{2} \Delta_x(T, \theta)^4 \mathcal{N}_{a,{\rm PBF}}^P  \frac{ \left(\frac{\omega}{2\Delta_x(T, \theta)} \right)^3}{\sqrt{\left(\frac{\omega}{2\Delta_x(T, \theta)} \right)^2-1}} \left[ f_F \left( \frac{\omega}{2T}\right) \right]^2 \,,
\end{align}
where $\mathcal{N}_{a,{\rm PBF}}^P$ is the normalization constant defined by $\int_{2\Delta (T, \theta)}^\infty J_{a,{\rm PBF}}^P d\omega = \epsilon_{a,{\rm PBF}}^P$; then $\mathcal{N}_{a,{\rm PBF}}^P = \epsilon_{a,{\rm PBF}}^P/T^5 I^{P}_{an} (z_{x0})$ with $z_{x0} \equiv \Delta_0^{(x)} (T)/T$.

\subsection{High-Energy Spectrum}
\label{SM:HE_spec}

\begin{figure}
\begin{center}
\includegraphics[width=0.49\linewidth]{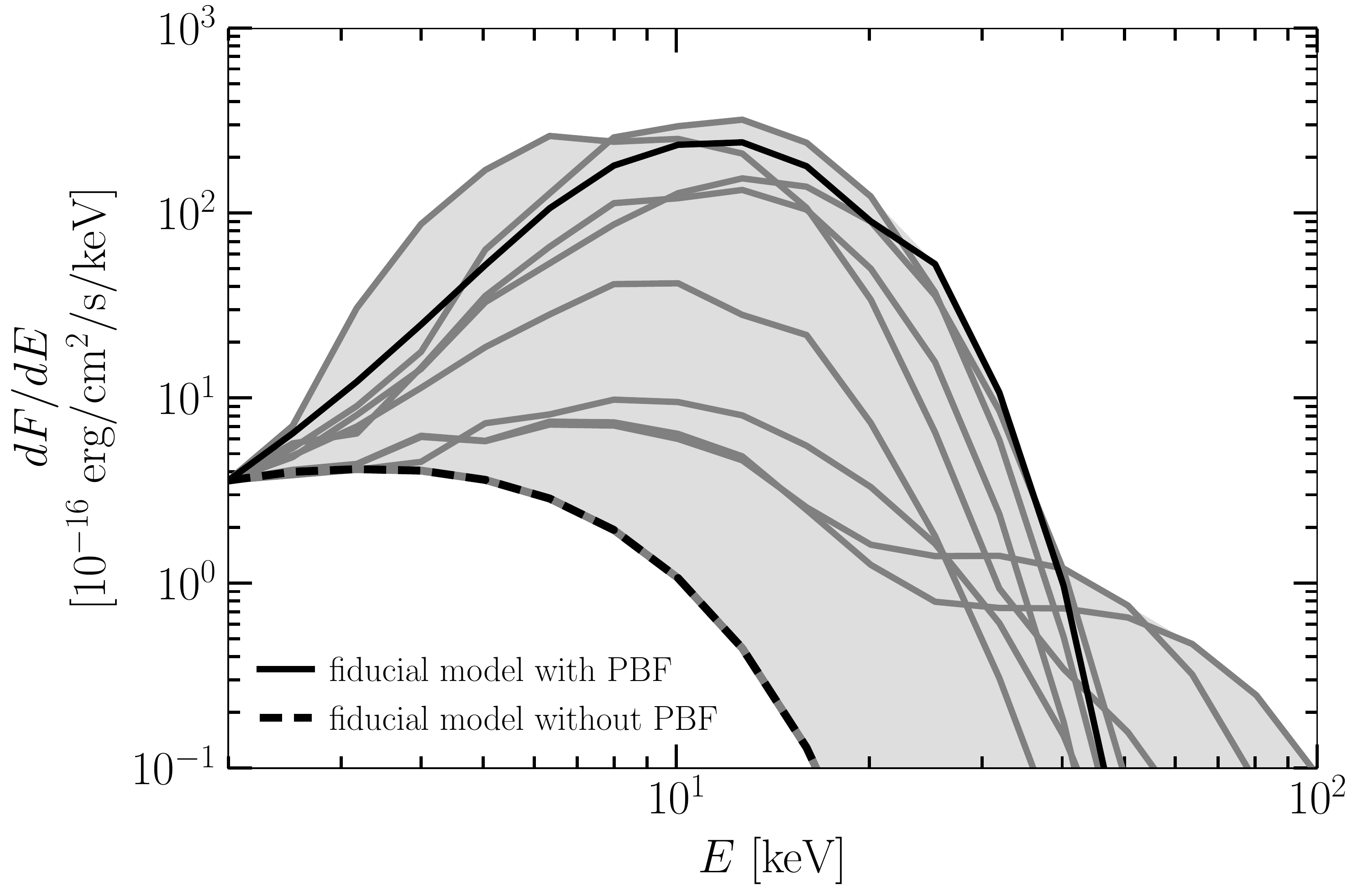}
\includegraphics[width=0.48\linewidth]{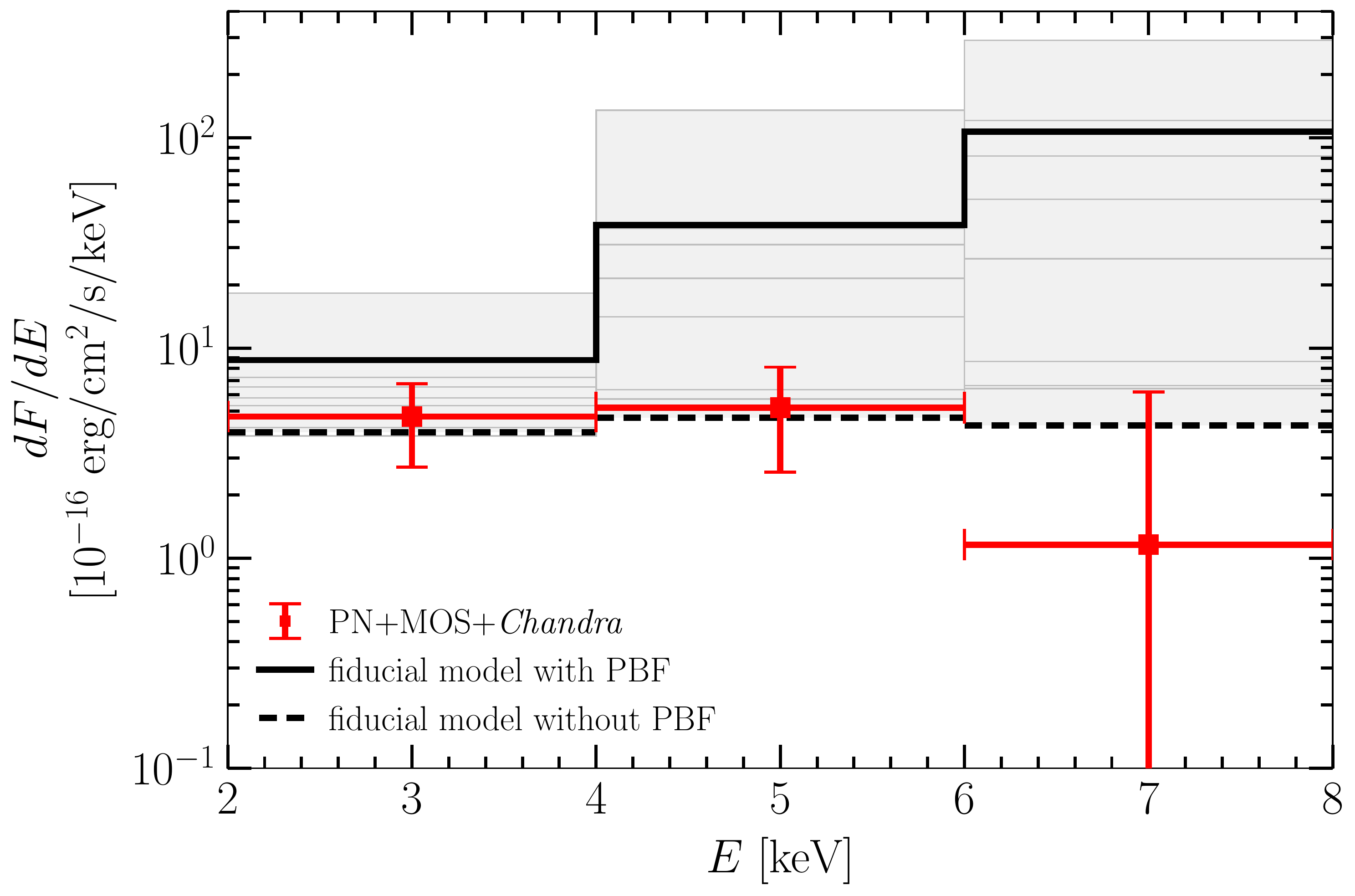}
\end{center}
\caption{(Left) The full energy spectrum from NS J1856 as predicted by the best fit of the axion model with the joint likelihood procedure performed in the main analysis at energies below 8 keV. The black dashed curve is the fiducial model we use in the main analysis, where no superfluidity is active, while the solid black curve shows the spectrum corresponding to our fiducial superfluidity model. The gray curves show the predictions from other superfluid models that we tested and the gray shaded region demonstrates the uncertainty as a result of the different superfluid models.  Note that there are three models that predict no enhancement and are overlapped with the black dashed curve. (Right) As in the left panel, but zoomed in below 8 keV and binned in 2 keV energy bins to provide a direct comparison to the $X$-ray data, which is also shown.}
\label{fig:PBF_spec_ave}
\end{figure}

At high energies, the flux may be dominated by the axions emitted in the PBF processes outlined in Sec.~\ref{SM:PBF_rates}. 
The spectral functions are sharply peaked at twice the gap energy, which is also the lower cutoff of the axion energy due to conservation of energy. The spectral functions then drop off quickly at higher energies. The exception is for the type B process in the $P$-wave pairing, where the gap energy $\Delta_B(T,\theta)$ is anisotropic and can be small when the neutron momentum is approximately aligned with the quantization axis. This implies that the energy of the axion is distributed to values lower than the magnitude of the gap energy $\Delta_0^B(T)$ and is thus not subject to a specific lower cutoff. This is to be contrasted with $\Delta(T)$ and $\Delta_A(T,\theta)$ for the S-wave pairing and type A P-wave pairing, where a sharp lower cutoff is present for a given $T$.

We show in Fig.~\ref{fig:PBF_spec_ave} the predicted spectrum at high energies for J1856 assuming the best-fit core temperature from the global axion model fit. Different curves denote different models~\cite{Hoffberg:1970vqj, Amundsen:1984qc, 1972PThPh..48.1517T, Baldo:1992kzz, Elgaroy:1996mg} used in \texttt{NSCool} for computing the superfluid critical temperatures of the NSs. Out of the twelve models available in  \texttt{NSCool}, there exist three models~\cite{1972PThPh..48.1517T, Elgaroy:1996mg} that do not lead to superfluidity formation and thus the production is given by the nucleon bremsstrahlung processes \new{as in the main text, where we assume no superfluidity}, and the predicted spectrum is given by the black dashed curve in Fig.~\ref{fig:PBF_spec_ave} in this case. With the black solid curve we show the spectrum from \new{our fiducial superfluidity model (model I) which includes PBF emission due to superfluidity formation.
That model takes the $^1S_0$ neutron pairing gap from~\cite{Schwenk:2002fq}, the neutron $^3P_2$ - $^3F_2$ pairing gap from ``model a'' in~\cite{Page:2004fy}, and the proton $^1S_0$ pairing gap from~\cite{BALDO1992349}, respectively, and predicts the maximum net high-energy intensity of all the superfluidity models considered.
} The gray shaded region spans between the black dashed curve and the maximum flux at each energy among the twelve models we scan over, representing an estimate of the model uncertainty in the flux. We note that we normalize the spectra of all models such that they all give the same value at 2 keV.  Note that it may be seen that even below $\sim$3 keV there are small deviations away from the spectrum assumed in the main text for some superfluid models due to the Type B $P$-wave superfluid pairing PBF process.

If we instead fix $g_{a\gamma\gamma} g_{ann} = 1 \times 10^{-20}$ GeV$^{-1}$ and take vanishing $m_a$, 
the predicted flux at 2 keV ranges from $2 \times 10^{-16}$ to $8 \times 10^{-15}$ erg/cm$^2$/s/keV for this NS depending on the superfluid model.  This shows that the superfluid model can significantly affect the low-energy flux as well due to the superfluid suppression factors, though these are energy independent and do not modify the spectral shape. 

We assumed $g_{app} = g_{ann}$ in Fig.~\ref{fig:PBF_spec_ave}, but the PBF flux is dominantly from the $P$-wave processes, which only involve neutrons. Thus we expect $g_{app}$ to play a less important role in the high-energy spectrum unless $g_{ann} \ll g_{app}$. Among the $P$-wave processes, emission from type B pairing dominates over that of type A. The predicted spectral shape is also highly dependent on the core temperature. At higher core temperatures, the spectral peak shifts to a higher energy. 

The M7 have not been studied in detail before at energies greater than $10$ keV. However, there are existing constraints from hard $X$-ray telescopes which we summarize now. The strongest constraint at these energies comes from the 105 month {\it Swift} Burst Alert Telescope all-sky hard $X$-ray survey~\cite{Oh:2018wzc}, which covers the full sky with median sensitivity $7 \times 10^{-12}$ erg/cm$^2$/s at $5\sigma$ in the 14 - 195 keV band. The predicted $X$-ray intensity from J1856 in this band assuming the fiducial model with \new{superfluidity} is $2 \times 10^{-13}$ erg/cm$^2$/s with a contribution from nucleon bremsstrahlung of $6 \times 10^{-16}$ erg/cm$^2$/s. For NSs near the galactic plane $|b| \leq 17.5^\circ$ (J1856, J0806, J0720, and J2143), constraints from the 14-year {\it INTEGRAL} galactic plane survey~\cite{Krivonos:2017hpc} with the IBIS camera apply. 90\% of the survey area is covered down to a 17 - 60 keV flux limit of $1.3 \times 10^{-11}$ erg/cm$^2$/s at $4.7\sigma$. Our fiducial \new{superfluidity} model predicts an intensity for J1856 in this range of $2 \times 10^{-13}$ erg/cm$^2$/s with a contribution from nucleon bremsstrahlung of $3 \times 10^{-16}$ erg/cm$^2$/s. This information is summarized in Tab.~\ref{tab:predicted_intensity}, where our fiducial model with \new{superfluidity} is denoted ``maximum'' and the nucleon bremsstrahlung contribution considered in the main text is denoted ``minimum''. Note, however, that the above limits assume a power-law intensity that peaks at low energies, whereas the axion intensity peaks at higher energies where both telescope effective areas are low, and the true limits on axion emission are likely weaker than reported here.

The {\it NuSTAR} telescope would currently provide the most sensitive search for ultra-hard $X$-ray emission ($\gtrsim$10 keV) from the M7. To date, {\it NuSTAR} has not observed any of the M7. In Tab.~\ref{tab:predicted_intensity} we show the projected sensitivity at $95\%$ confidence for a 400 ks {\it NuSTAR} observation of J1856 in two energy bands, along with the predicted intensities in each model. $t_{\rm exp}$ = 400 ks is a comparable total exposure time to the {\it XMM} and {\it Chandra} exposure times for the M7~\cite{dessert2019hard}, and would confirm the emission below 10 keV and constrain the emission above 10 keV in some \new{superfluidity} models. Some of the models with \new{superfluidity} can be ruled out or confirmed with only a few ks of observation time.

\begingroup
\setlength{\tabcolsep}{4pt} 
\renewcommand{\arraystretch}{1.3} 
\begin{table}[]
\begin{tabular}{|rr||c||c||c|c|}
\hline
\multicolumn{2}{|c||}{\multirow{2}{*}{Energy Range}} & \multirow{2}{*}{Current Limit} & Projected Sensitivity & \multicolumn{2}{c |}{Predicted Intensity} \\ \cline{5-6} 
  &  & & ($t_{\rm exp} = $ 400 ks) & minimum & maximum  \\ \hline
{\it Swift}:  &  14 - 195 keV & 		$7 \times 10^{-12}$ 		&	--- 	& 	 $6 \times 10^{-16}$ 		&	$2 \times 10^{-13}$  \\ \hline
{\it INTEGRAL}:  &  17 - 60 keV &		 $1.3 \times 10^{-11}$ 		&	---	&	 $3 \times 10^{-16}$		&	$2 \times 10^{-13}$  \\ \hline
{\it NuSTAR}:  &  6 - 10 keV &		 --- 		&	$3 \times 10^{-15}$	&	 $2 \times 10^{-15}$		&	$6 \times 10^{-14}$  \\ \hline
{\it NuSTAR}:  &  10 - 60 keV &		 --- 		&	$2 \times 10^{-14}$	&	 $2 \times 10^{-15}$		&	$3 \times 10^{-13}$  \\ \hline
\end{tabular}
\label{tab:predicted_intensity}
\caption{The second and third columns show the current limit and future sensitivity on the $X$-ray intensity, whereas the last two columns list the maximum and minimum intensities predicted among the different superfluid models assuming the best fit of the axion model to the J1856 joint data. The ``maximum" predicted intensity assumes \new{our fiducial superfluidity} model, which predicts the largest intensity in 10 - 60 keV band. The ``minimum'' predicted intensity is the nucleon bremsstrahlung contribution discussed in the main text. All intensities are in units of erg/cm$^2$/s.
}
\end{table}
\endgroup

\section{Systematic Tests}
\label{SM:sysTest}

In this SM section we consider multiple systematic variations to the analysis procedure presented in the main text.  We begin by considering the consistency of the axion model between the three different cameras to assess possible systematic effects that only affect individual cameras.  In the next subsection we restrict and broaden the energy range relative to our fiducial analysis to analyze the robustness of the signal to changes in the energy range used in the analysis.  Next, we analyze separately the NSs that observe an excess and those that do not to more quantitatively address the consistency between the null results and detections.  In the following subsection we relax the restriction that $g_{app} = g_{ann}$ in the fit of the axion model to the $X$-ray data. We then consider how the best-fit axion parameter space and upper limit depend on the superfluidity model.  Lastly, we consider alternate models for the NS magnetic field strengths and surface temperatures.

\subsection{Dependence on instrument}

\begin{figure}[!htb]
    \centering
    \begin{tabular}[t]{cc}
\begin{minipage}{0.49\textwidth}
    \centering
    \smallskip
    \includegraphics[width=\linewidth]{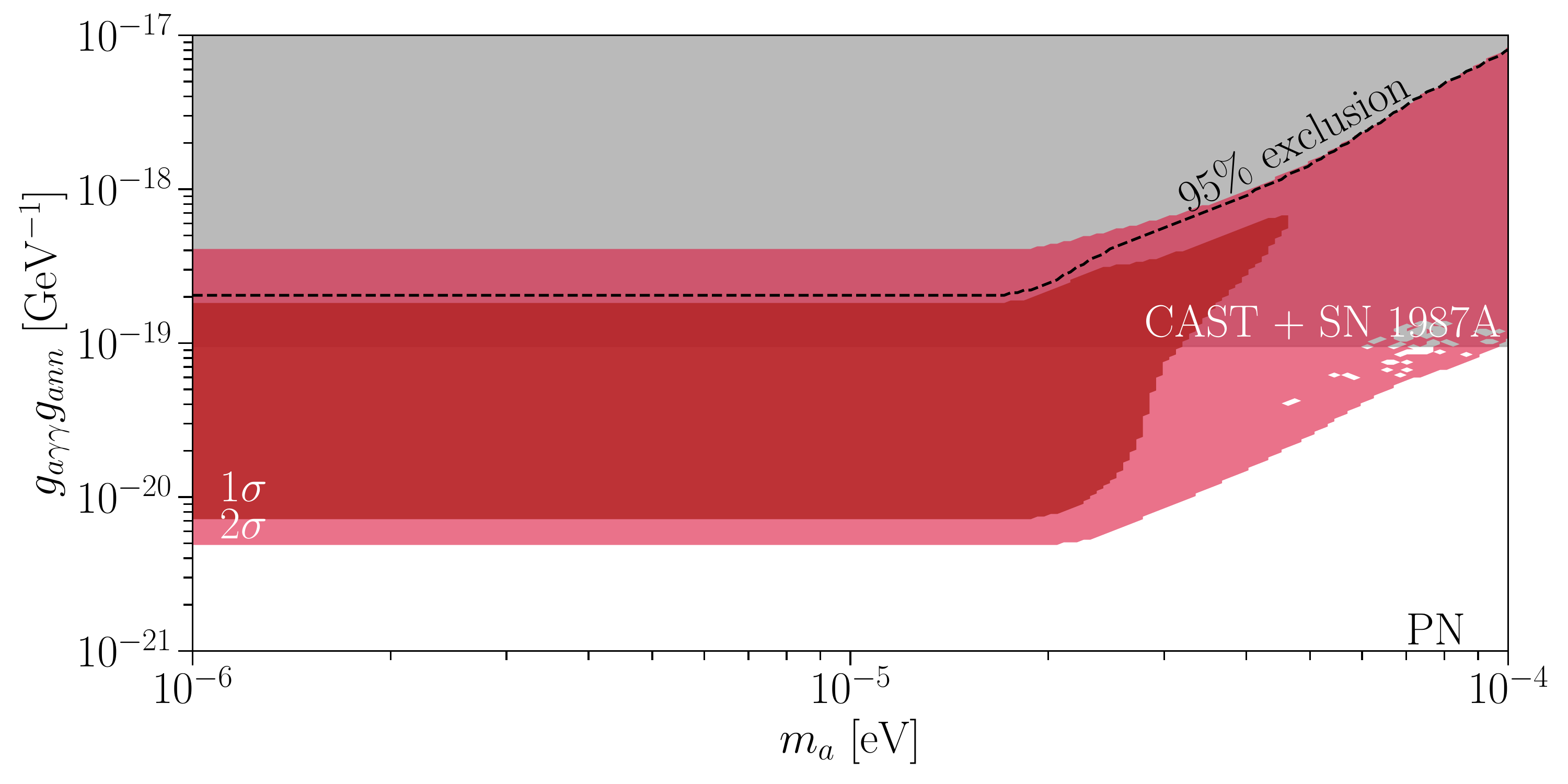}
\end{minipage}
    &
        \begin{tabular}{c}
        \smallskip
            \begin{minipage}[t]{0.49\textwidth}
                \centering
                \includegraphics[width=\textwidth]{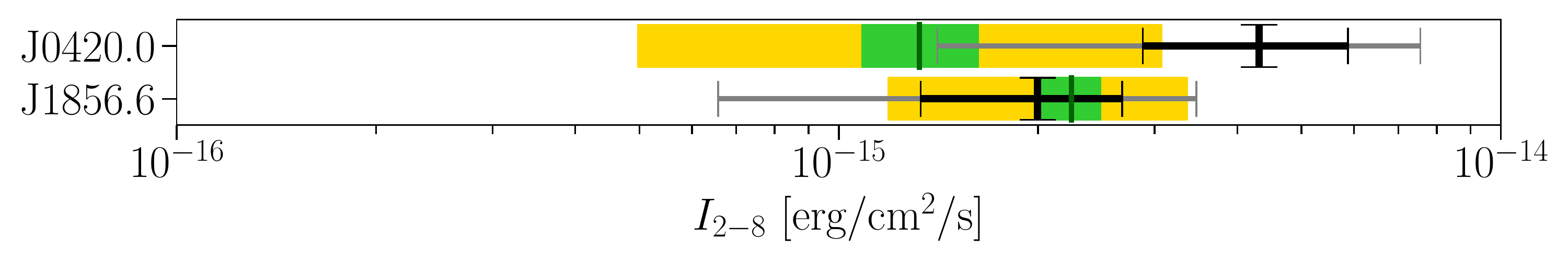}
            \end{minipage}\\
            \begin{minipage}[t]{0.49\textwidth}
                \centering
                \includegraphics[width=\textwidth]{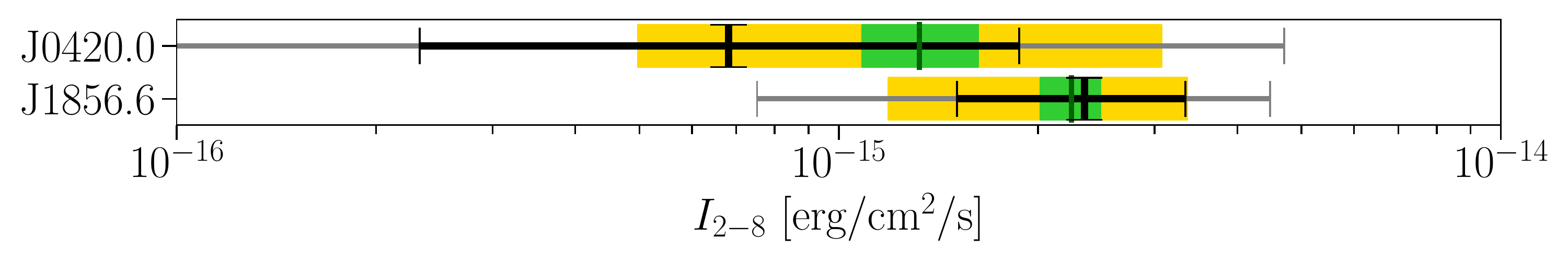}
            \end{minipage} \\
            \begin{minipage}[t]{0.49\textwidth}
                \centering
                \includegraphics[width=\textwidth]{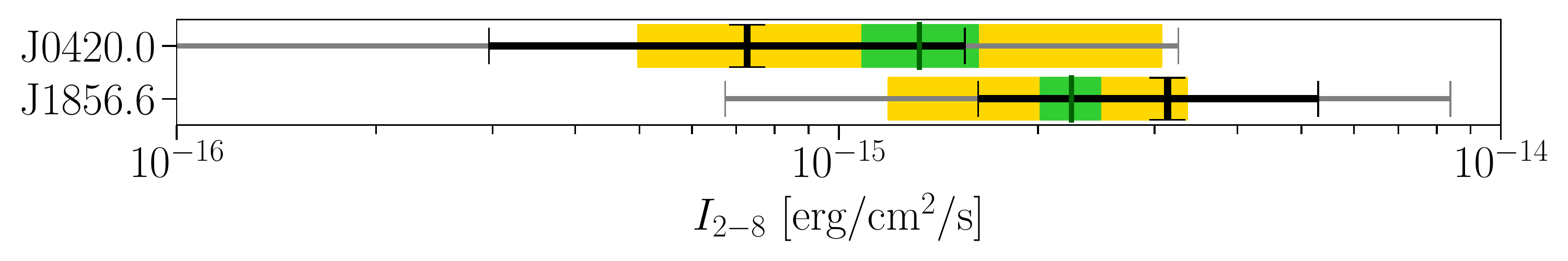}
            \end{minipage}
        \end{tabular}\\ \vspace{0.1cm}
    \end{tabular}
        \begin{tabular}[t]{cc}
\begin{minipage}{0.49\textwidth}
    \centering
    \smallskip
    \includegraphics[width=\linewidth]{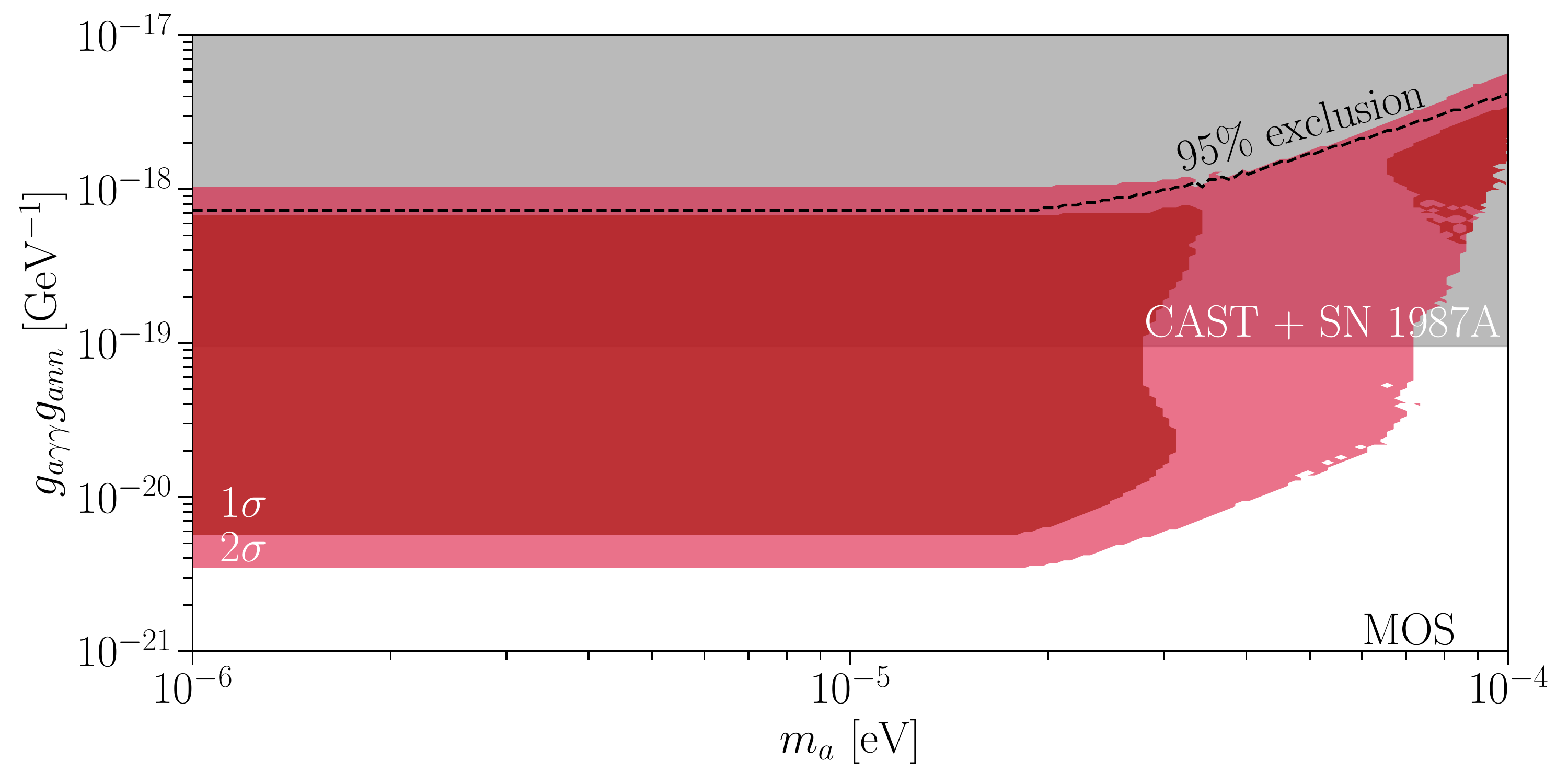}
\end{minipage}
    &
        \begin{tabular}{c}
        \smallskip
            \begin{minipage}[t]{0.49\textwidth}
                \centering
                \includegraphics[width=\textwidth]{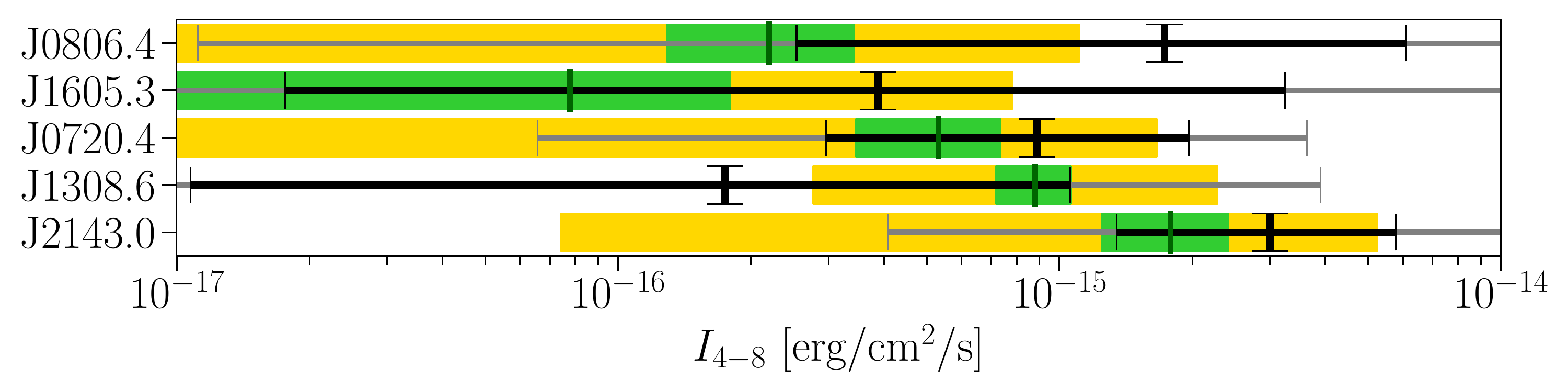}
            \end{minipage}\\
            \begin{minipage}[t]{0.49\textwidth}
                \centering
                \includegraphics[width=\textwidth]{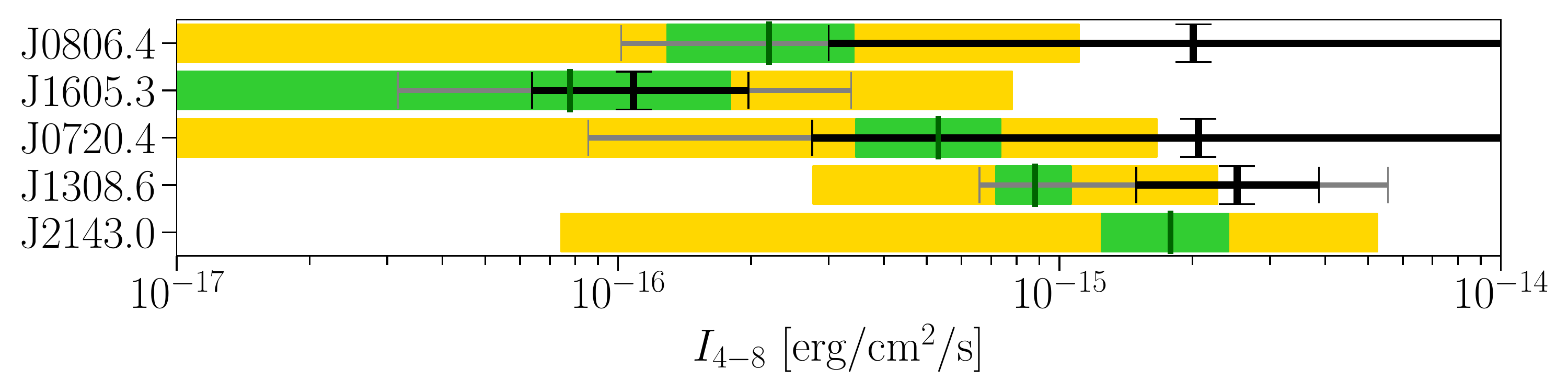}
            \end{minipage} \\
            \begin{minipage}[t]{0.49\textwidth}
                \centering
                \includegraphics[width=\textwidth]{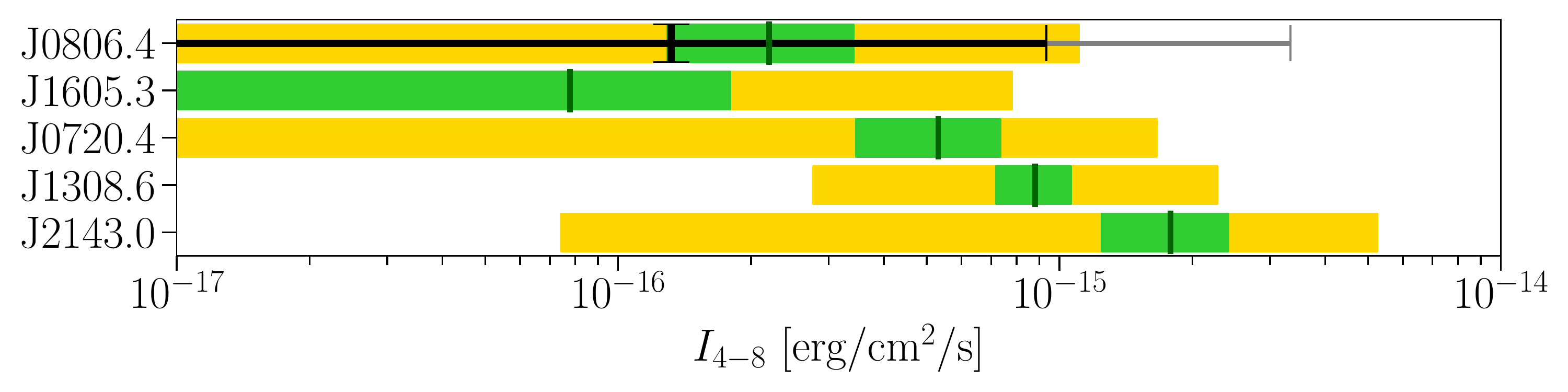}
            \end{minipage}
        \end{tabular} \\ \vspace{0.1cm}
    \end{tabular}
        \begin{tabular}[t]{cc}
\begin{minipage}{0.49\textwidth}
    \centering
    \smallskip
    \includegraphics[width=\linewidth]{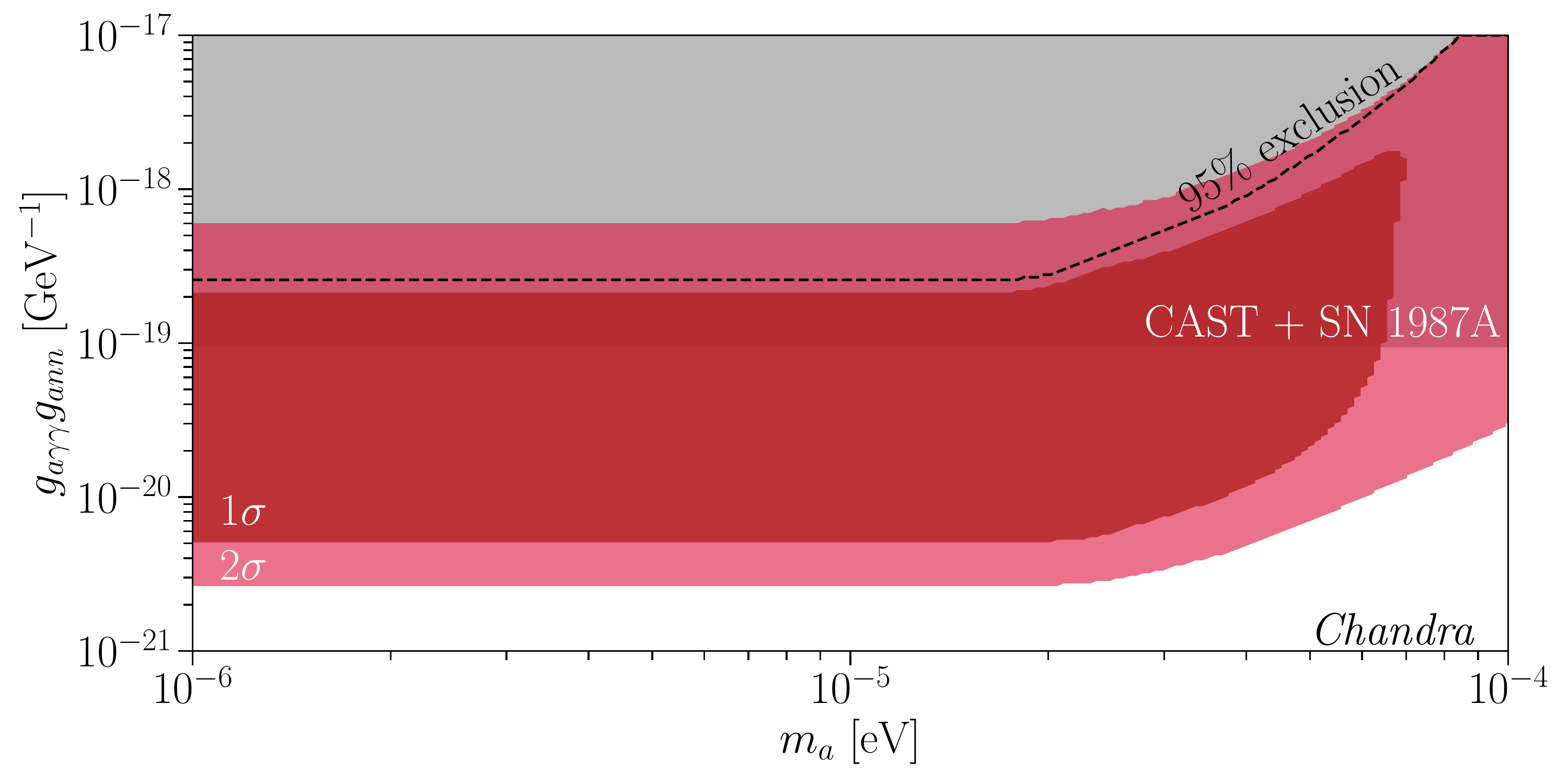}
\end{minipage}
    &
        \begin{tabular}{c}
        \smallskip
            \begin{minipage}[t]{0.49\textwidth}
                \centering
                \includegraphics[width=\textwidth]{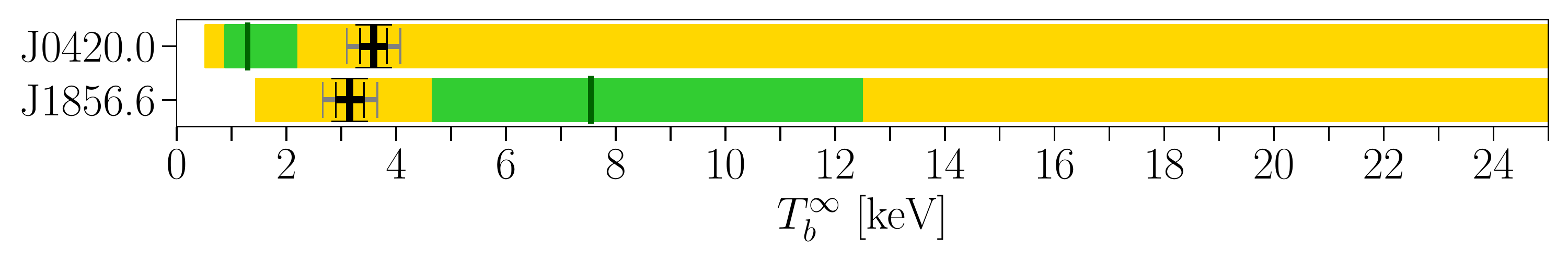}
            \end{minipage}\\
            \begin{minipage}[t]{0.49\textwidth}
                \centering
                \includegraphics[width=\textwidth]{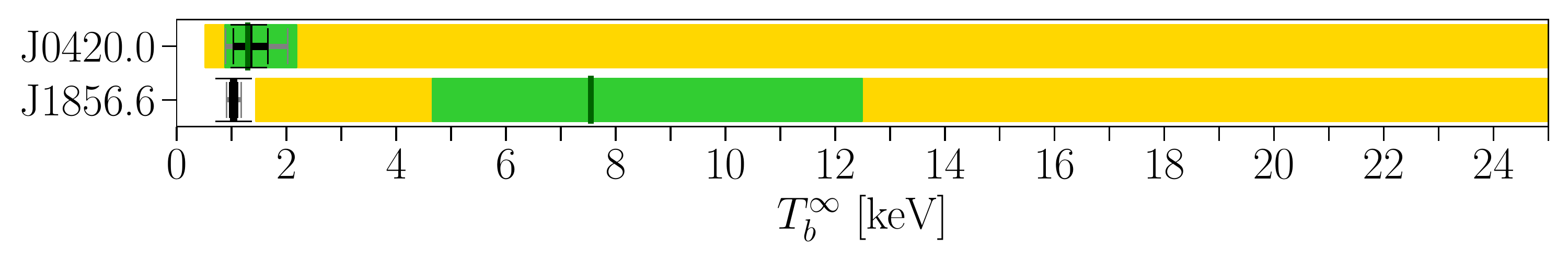}
            \end{minipage} \\
            \begin{minipage}[t]{0.49\textwidth}
                \centering
                \includegraphics[width=\textwidth]{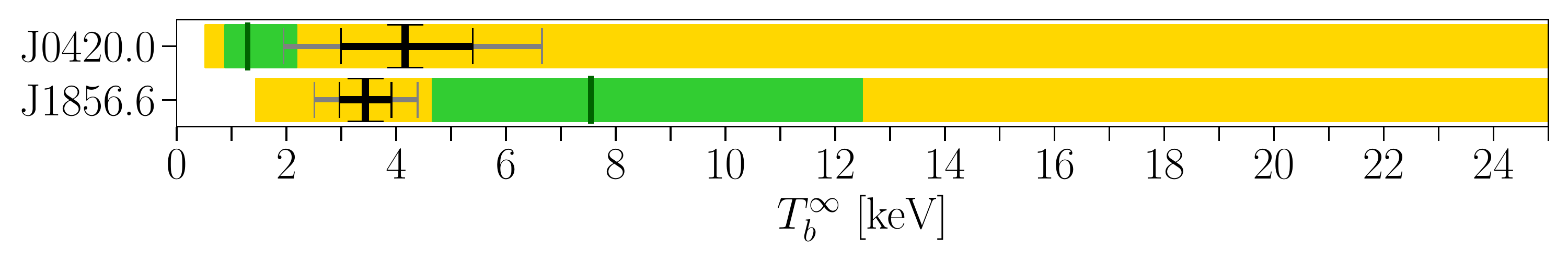}
            \end{minipage}
        \end{tabular}\\
    \end{tabular}
    \caption{As in Figs.~\ref{fig:money},~\ref{fig:I28}, and~\ref{fig:Ts} except combining the data from PN, MOS, and {\it Chandra} separately, as indicted.  We find non-trivial and consistent evidence for the axion model between datasets.}
    \label{fig:money-cameras}
\end{figure}

In~\cite{dessert2019hard} we show that all three cameras (PN, MOS, and {\it Chandra}) give consistent spectra for the M7 hard $X$-ray flux.  This is highly non-trivial considering that these instruments respond differently to {\it e.g.}~pileup and unresolved point sources.  Given that the observed fluxes are consistent between the three cameras, we also expect the best-fit axion parameter space regions to be consistent between the different cameras.  Indeed, as we show in Fig.~\ref{fig:money-cameras}, we observe this to be the case.
In that figure we show the best-fit axion parameter space as in Fig.~\ref{fig:money} but determined using the data from each camera independently, as indicated.  Interestingly, we find significant evidence in favor of the axion model from each camera independently.  We also show the observed intensities $I_{2-8} (I_{4-8})$, as described in the main text, and the best-fit temperatures.  

\subsection{Dependence on the energy range}

\begin{figure}[htb]
\hspace{0pt}
\vspace{-0in}
\begin{center}
\includegraphics[width=0.49\textwidth]{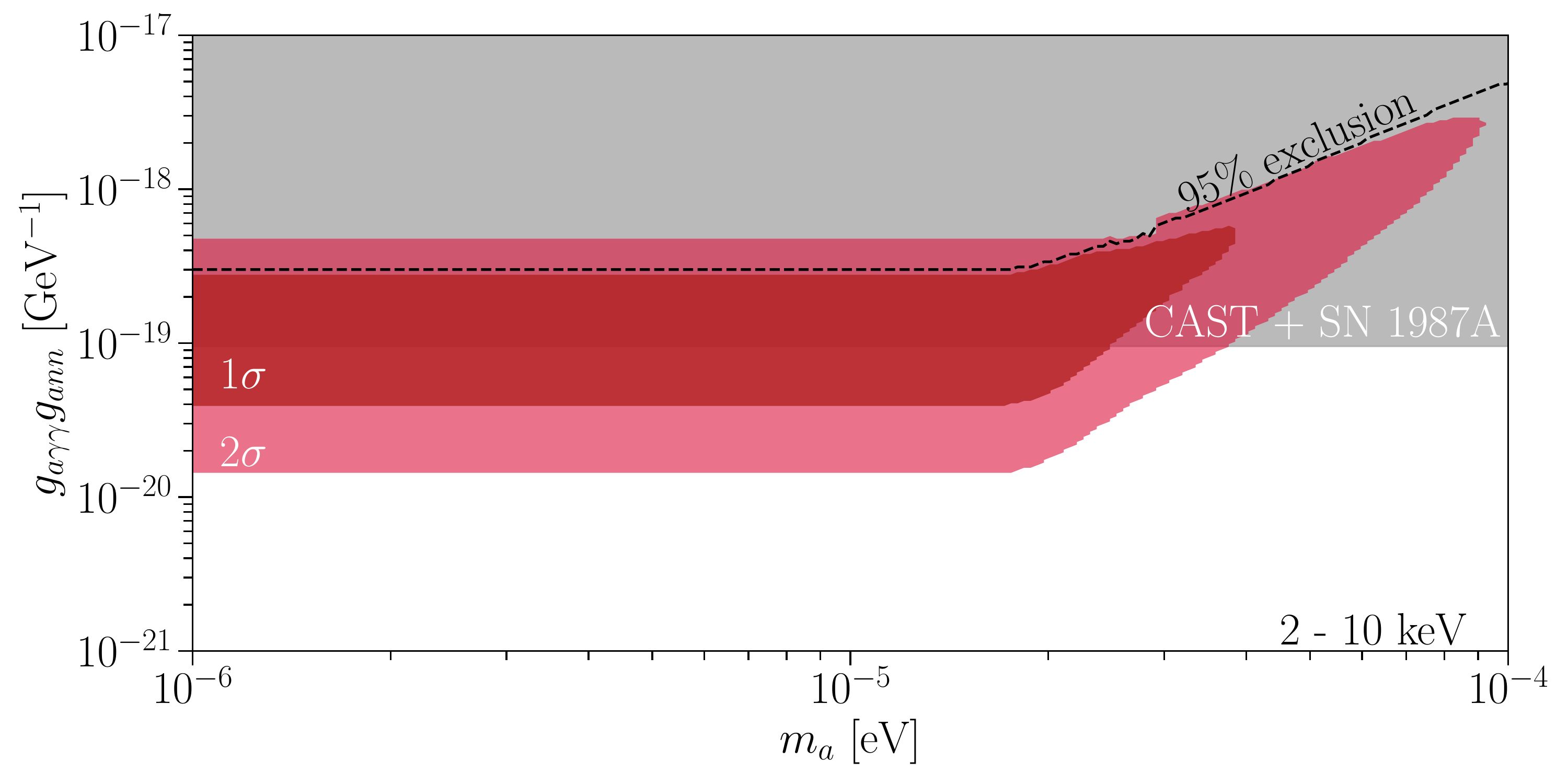} 
\includegraphics[width=0.49\textwidth]{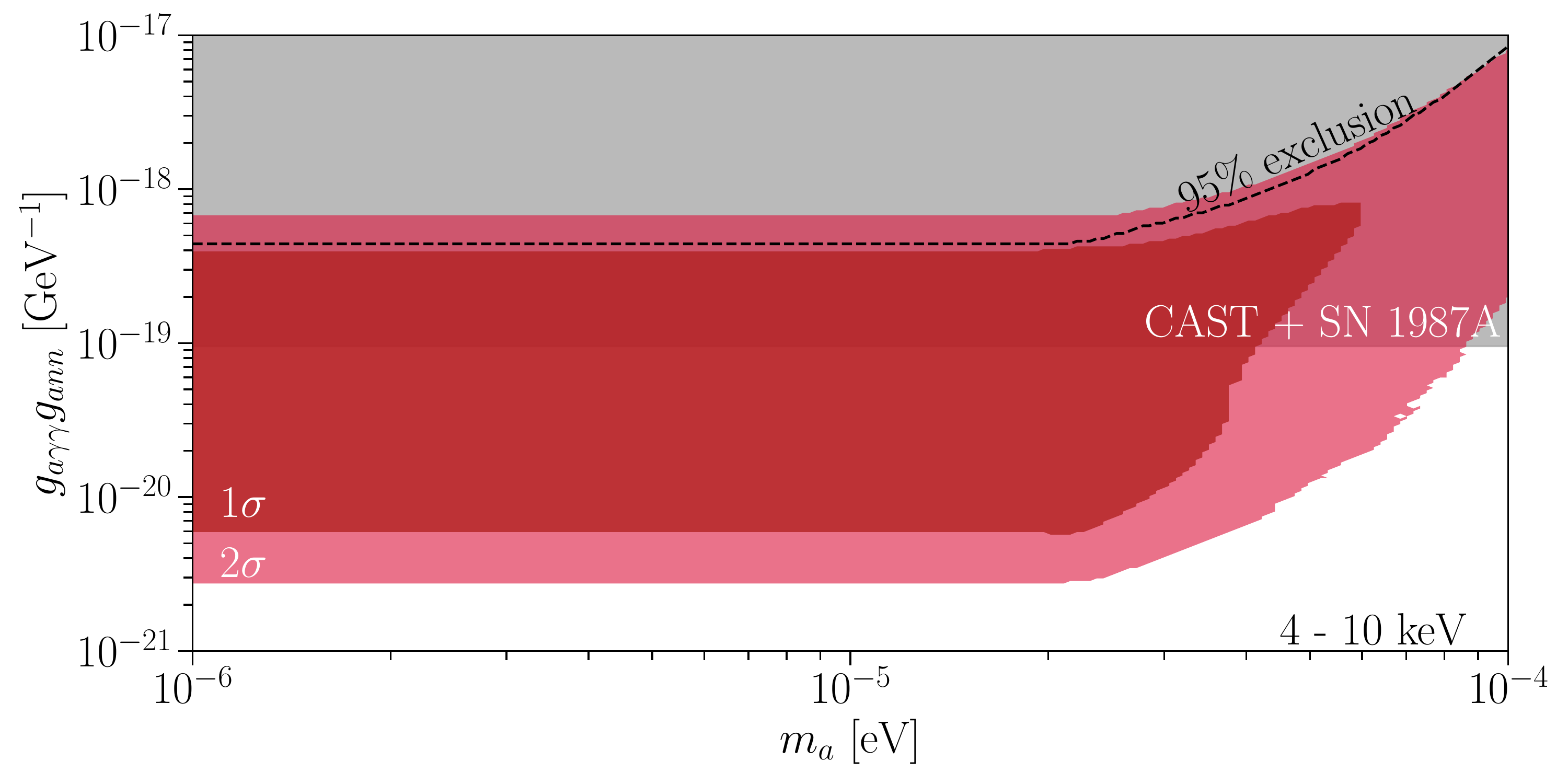} \\
\includegraphics[width=0.49\textwidth]{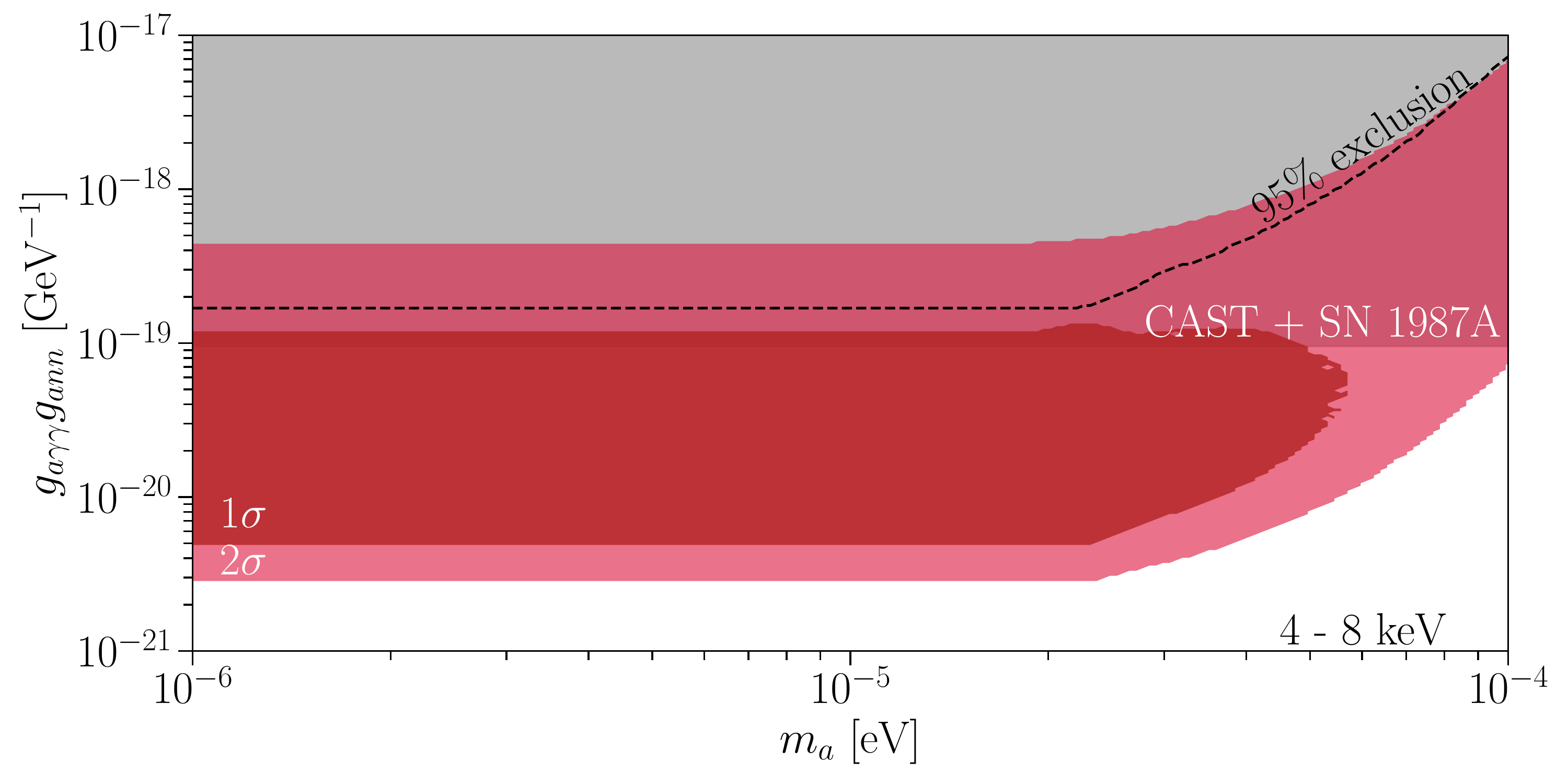}
\caption{
\label{fig:money-4-10}
As in Fig.~\ref{fig:money} and Fig.~\ref{fig:I28} but with variations to the choices of energy bins included in the analysis.  (Top Left) We use the fiducial energy bin choices plus the 8 - 10 keV bin for all NSs.  (Top Right) We use the energy bins 4 - 6 keV, 6 - 8 keV, and 8 - 10 keV for all NSs.  (Bottom) We use the energy bins 4 - 6 keV and 6 - 8 keV for all NSs.
}
\end{center}
\end{figure}

In the main text we used three energy bins from 2 - 4, 4 - 6, and 6 - 8 keV for J1856 and J0420, while for the other 5 NSs we only used the last two energy bins.  We find that removing the 2 - 4 keV energy bin for J1856 and J0420 leads to consistent results.  Additionally, in~\cite{dessert2019hard} data is also presented for the 8 - 10 keV energy bin.  For both {\it XMM-Newton} and {\it Chandra} this energy bin suffers from increased statistical and systematic uncertainties, as it is at the edge of the energy range of the cameras, so it is not included in the fiducial analysis.  Still, it is reassuring to see that including this energy bin does not substantially influence the global fit, which is mostly due to the fact that the uncertainties in that bin are quite large.  To emphasize these points in Fig.~\ref{fig:money-4-10} we show the best-fit axion parameter space and upper limit for variations to the energy bin choices.  In the top left panel we use our fiducial energy bin choice but add in the 8 - 10 keV bin for all NSs.  The top right panel is as in the top left but with the 2 - 4 keV bin removed for J1856 and J0420.  Lastly, the bottom panel is as in the top right but without the 8 - 10 keV bin.

\subsection{Influence of different neutron stars}

\begin{figure}[htb]
\hspace{0pt}
\vspace{-0in}
\begin{center}
\includegraphics[width=0.49\textwidth]{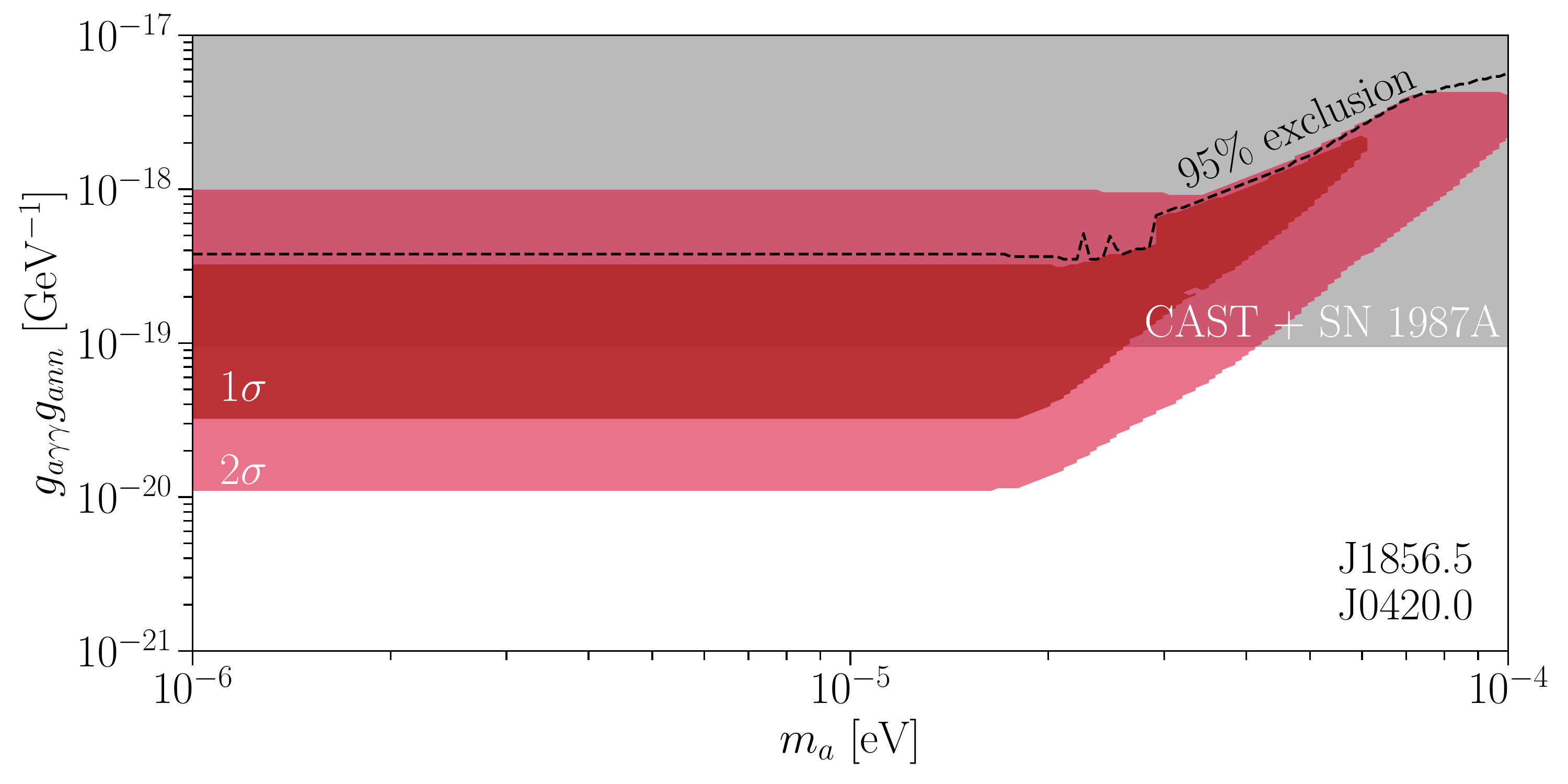}
\includegraphics[width=0.49\textwidth]{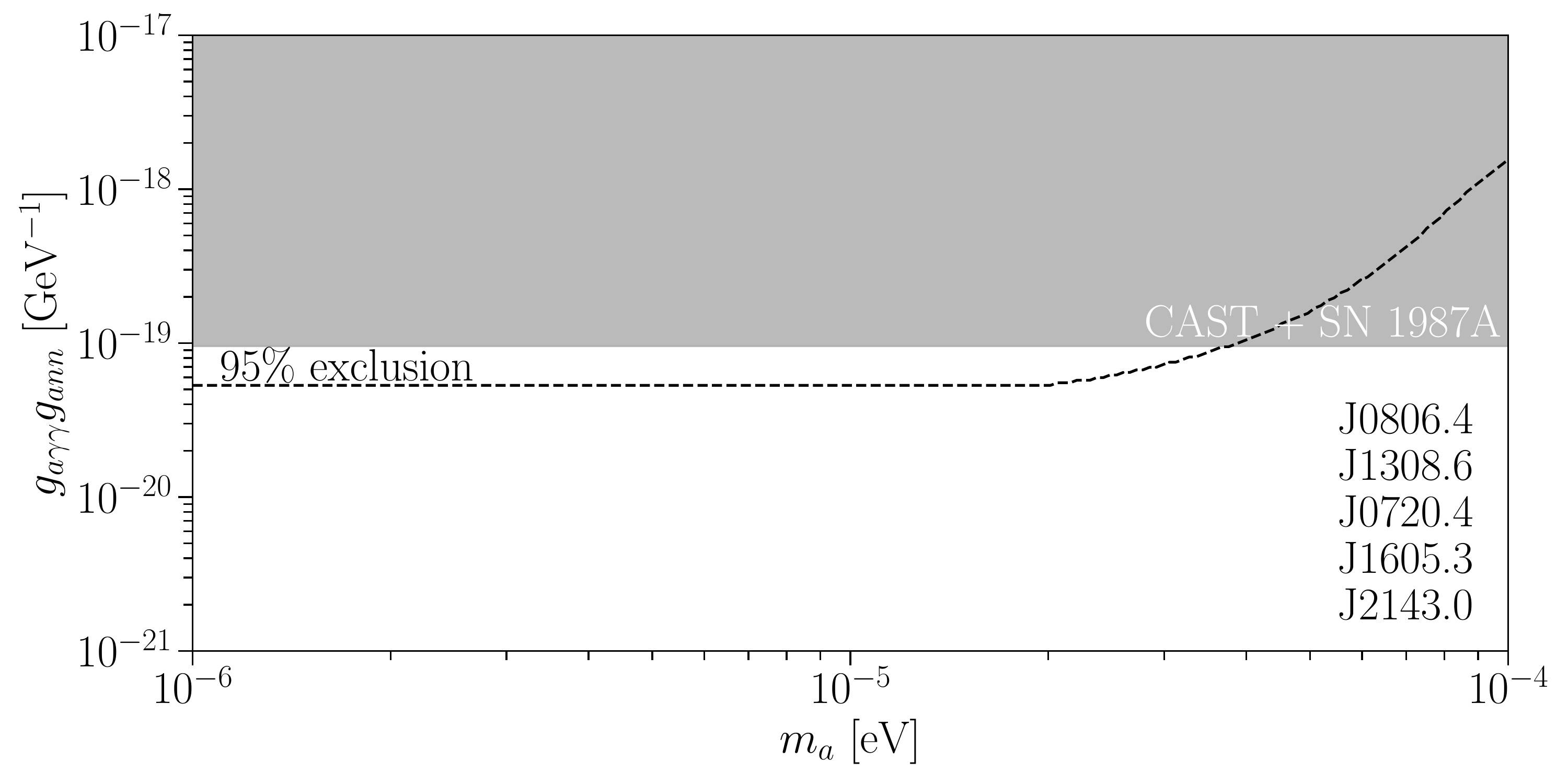}
\caption{
\label{fig:money-diff-NSs}
As in Fig.~\ref{fig:money} but (left) only including J1856 and J0420 and (right) only including the other five NSs. In the right panel we find less than $\sim$1$\sigma$ evidence (with two degrees of freedom) for the axion model when fitting to the other five neutron stars.  Only the 95\% upper limit is shown in this case.
}
\end{center}
\end{figure}

The evidence in favor of the axion model is driven the most by the high-significance excesses in the two NSs J1856 and J0420.  In Fig.~\ref{fig:money-diff-NSs} we perform a combined fit to the J1856 and J0420 data and then a separate combined fit to the data from the other five NSs.  In the second fit we find marginal evidence (slightly less than $\sim$1$\sigma$ with two degrees of freedom) for the axion model.    

\subsection{Dependence on the nucleon couplings}

\begin{figure}[htb]
\hspace{0pt}
\vspace{-0in}
\begin{center}
\includegraphics[width=0.49\textwidth]{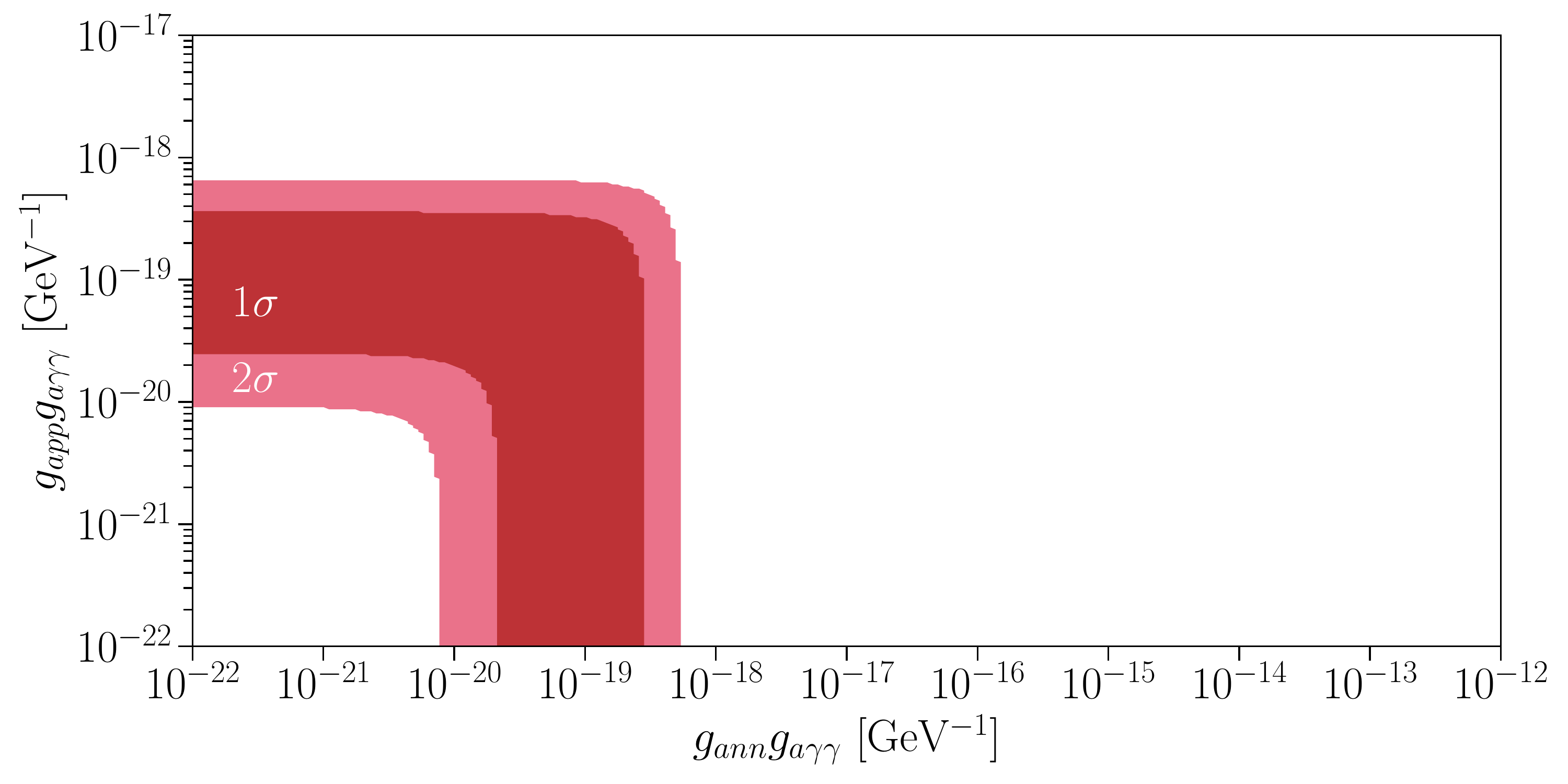}
\includegraphics[width=0.49\textwidth]{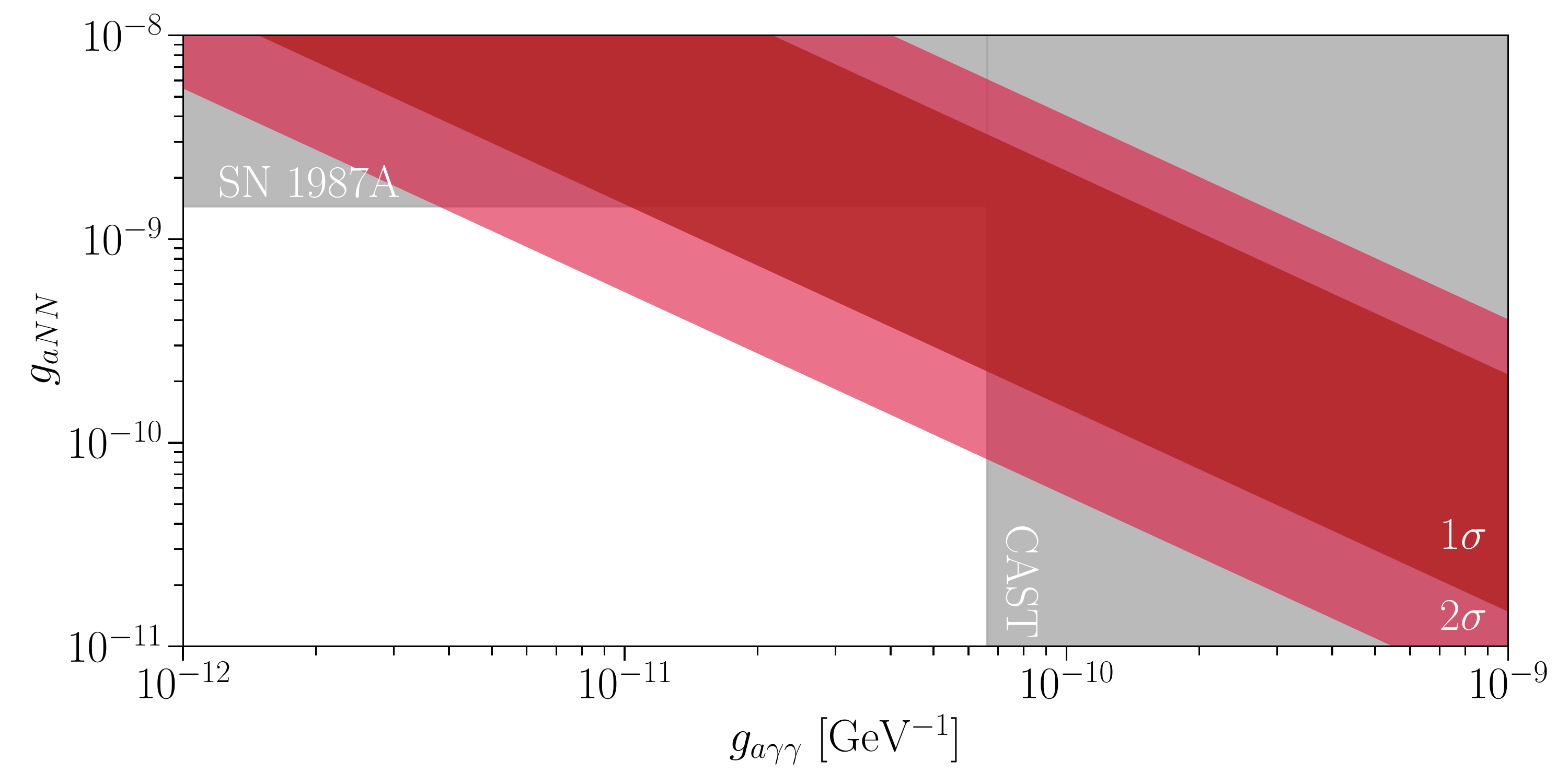}\\
\includegraphics[width=0.49\textwidth]{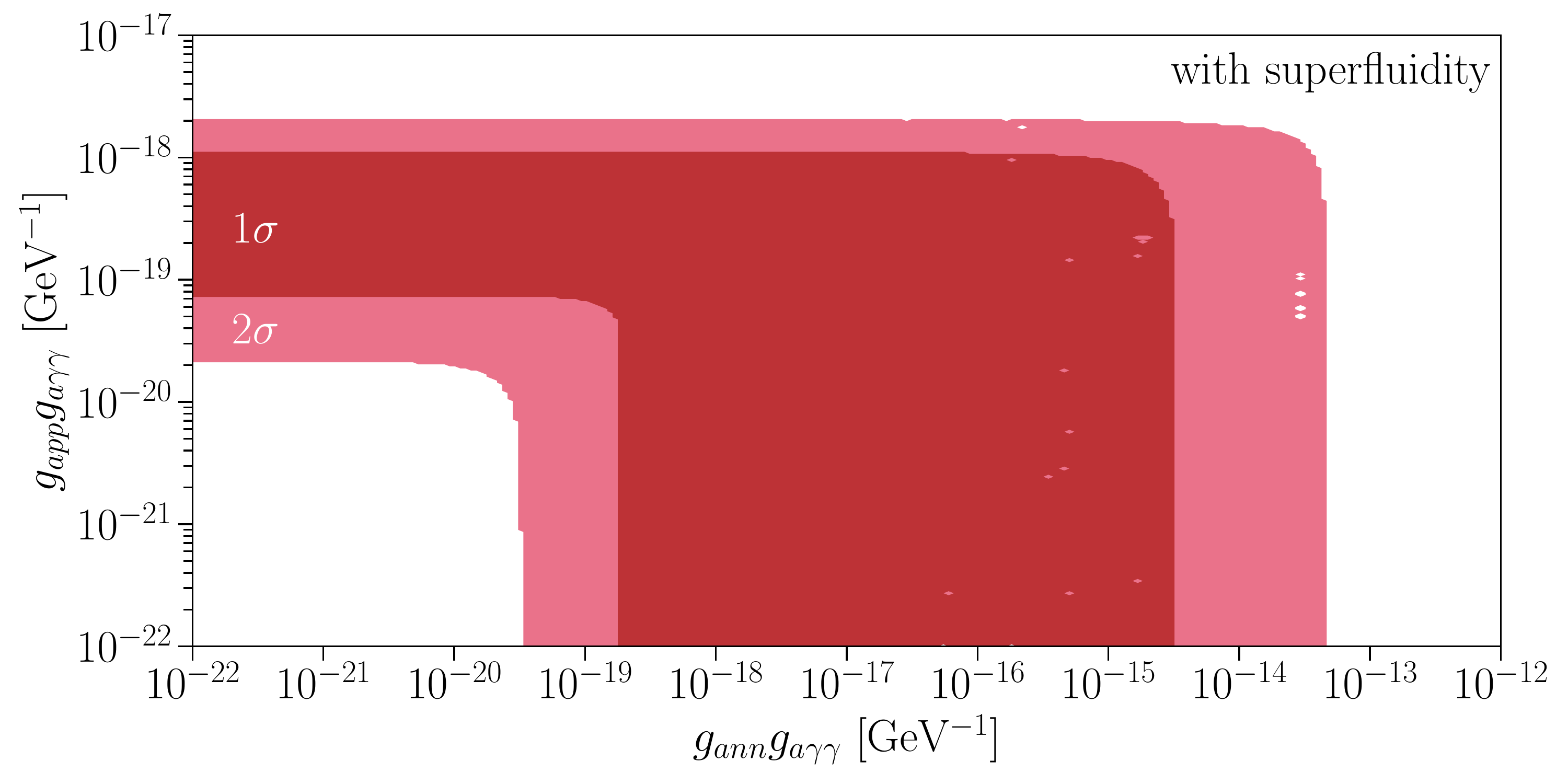}
\includegraphics[width=0.49\textwidth]{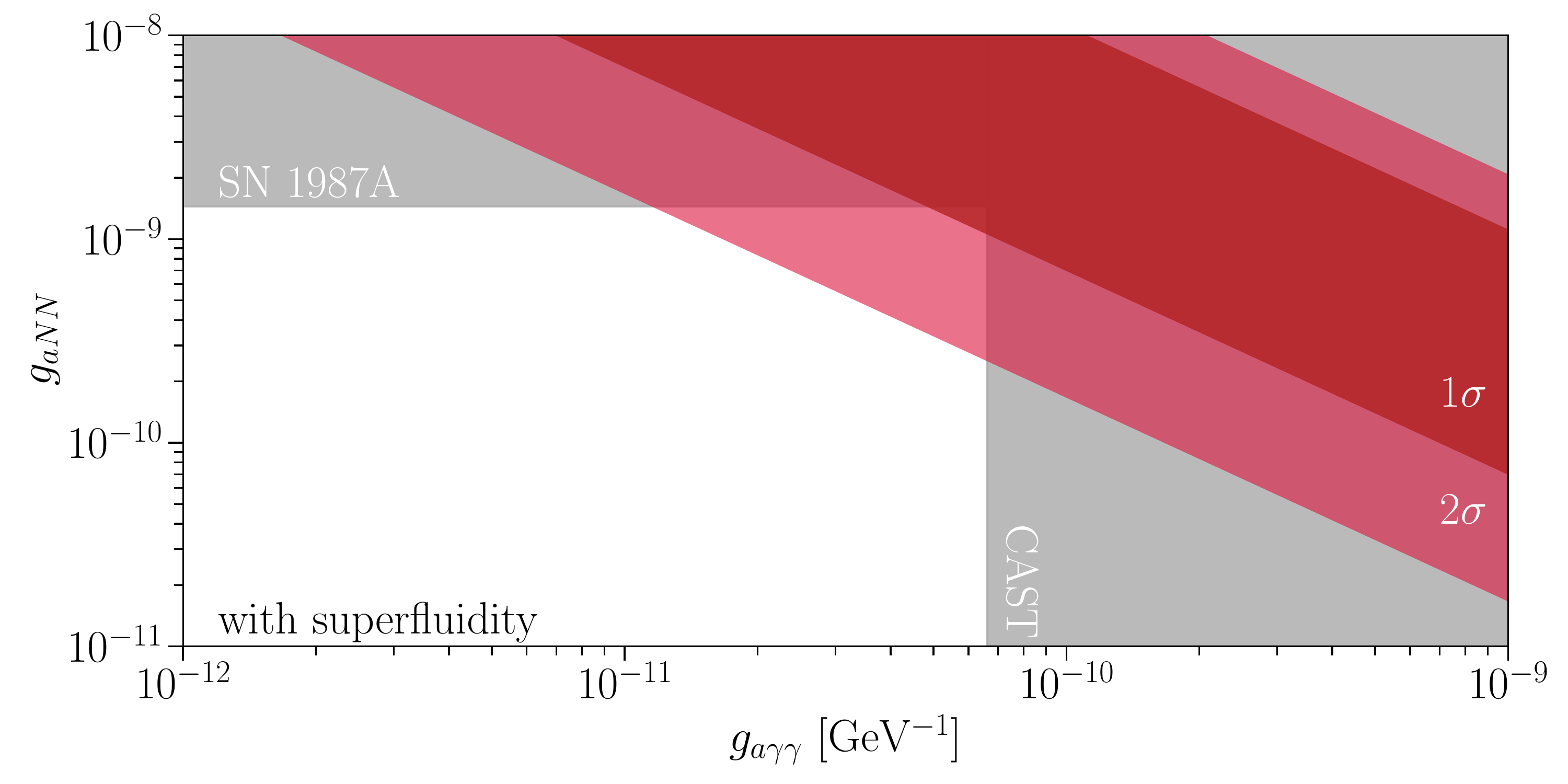}
\caption{
\label{fig:money-relax}
\new{Best fit 1 and 2$\sigma$ parameter space in our fiducial analysis under the assumption of vanishing axion mass ($m_a \ll 10^{-5}$ eV). In the left panels we relax the constraint $g_{app} = g_{ann}$ and in right panels we disentangle $g_{ann}$ (with $g_{app} = g_{ann}$ fixed) and $g_{a\gamma\gamma}$. The bottom row is the same as the top row but assuming nucleon superfluidity.} 
}
\end{center}
\end{figure}

In the main text we took, for definiteness, $g_{app} = g_{ann}$ in all figures.  In this section we relax that assumption under the condition of vanishing axion mass ($m_a \ll 10^{-5}$ eV).  In the left panels of Fig.~\ref{fig:money-relax} we show the best-fit axion model space in the $g_{a\gamma\gamma}g_{ann}$-$g_{a\gamma\gamma}g_{app}$ plane.  \new{Importantly, note that comparable neutron and proton axion couplings may lead to comparable $X$-ray fluxes.  However, with superfluidity included (bottom left panel, which uses our fiducial superfluidity model I) it is possible that the neutron axion production mechanism is significantly suppressed relative to that from the proton, since the neutron superfluid transition temperature is generically higher than that of the proton in this model.  The axion production rates are exponentially suppressed below the superfluid transition temperature, which requires higher axion couplings to produce the same $X$-ray flux.  In the bottom left panel we show the best-fit region without including superfuidity, as in the main text.  
In this case, the neutron and proton couplings produce comparable results.}

\new{
In the right panels of Fig.~\ref{fig:money-relax} we again take $m_ \ll 10^{-5}$ eV, but we fix $g_{ann} = g_{app}$ and we illustrate the best-fit region in the $g_{ann}-g_{a\gamma\gamma}$ plane.  The top panel does not include superfluidity while the bottom panel does.  To be consistent with current constraints from SN 1987A and CAST, the axion model should reside in the regions that are not shaded grey.  
}

\subsection{Dependence on superfluidity model}

\begin{figure}[htb]
\hspace{0pt}
\vspace{-0in}
\begin{center}
\includegraphics[width=0.49\textwidth]{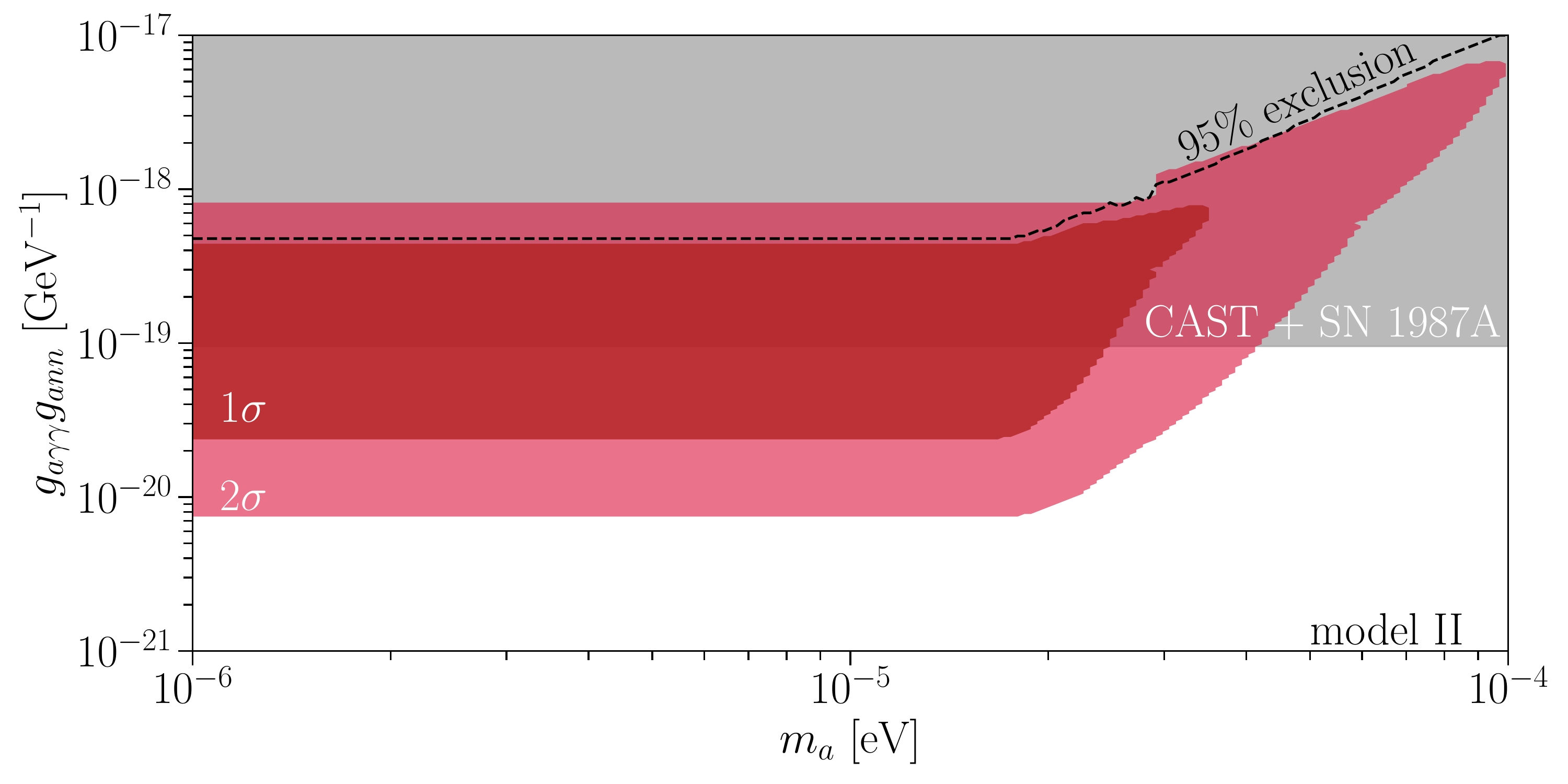}
\includegraphics[width=0.49\textwidth]{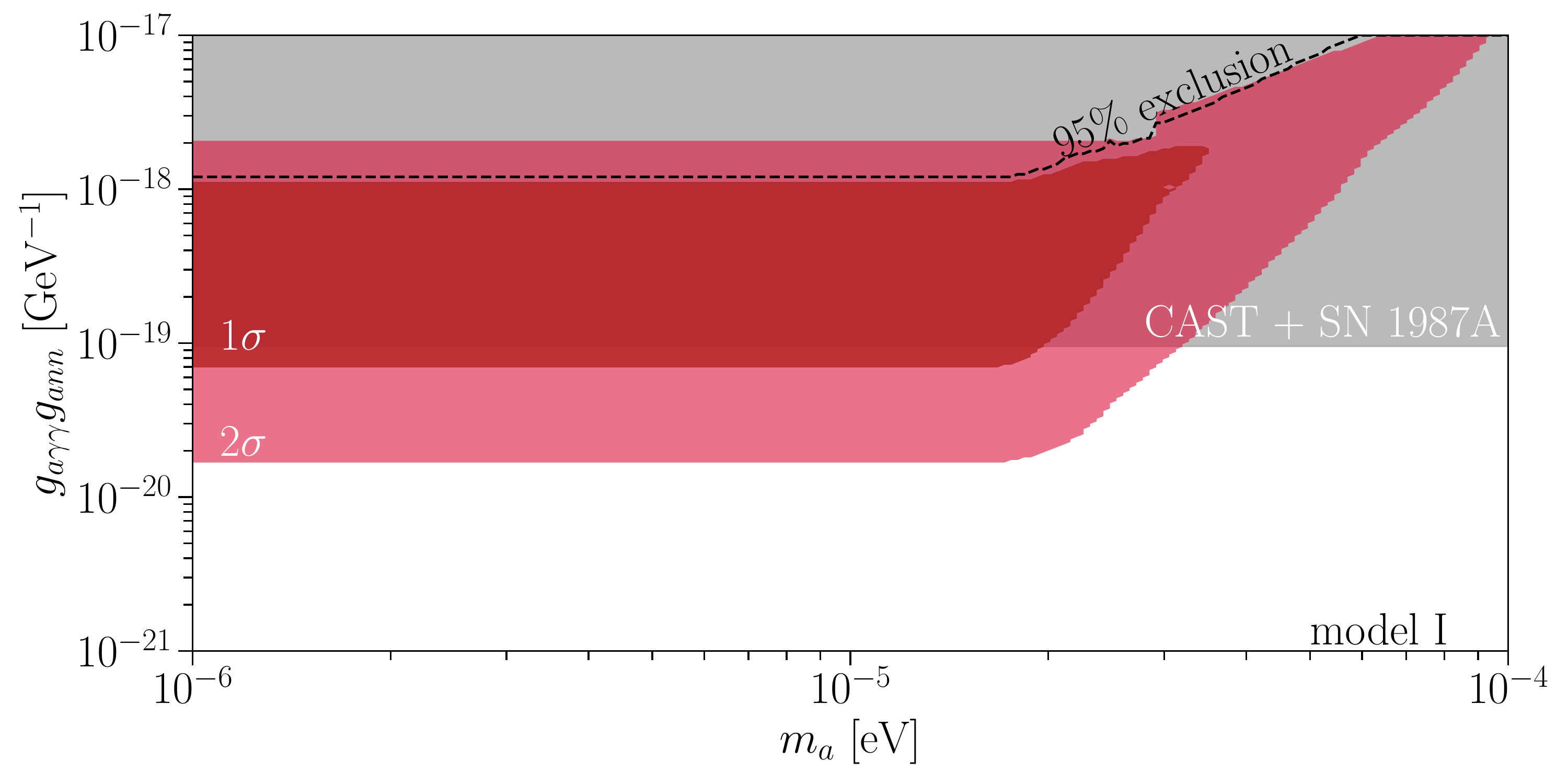}\\
\includegraphics[width=0.49\textwidth]{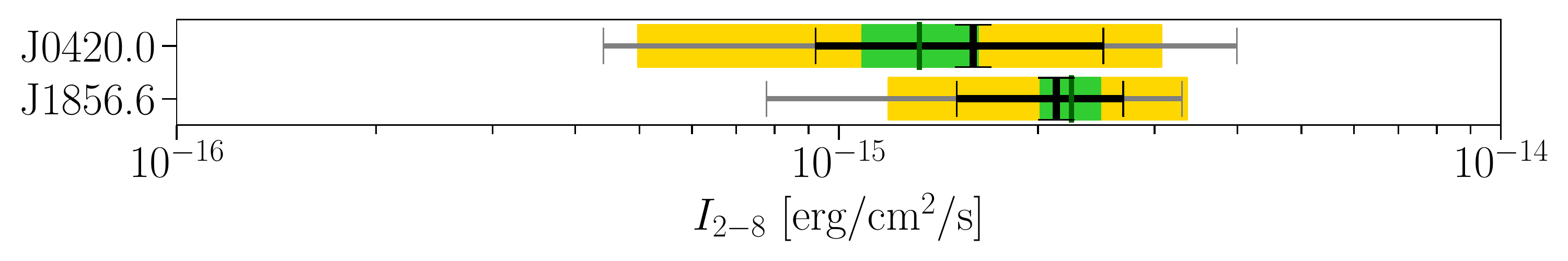}
\includegraphics[width=0.49\textwidth]{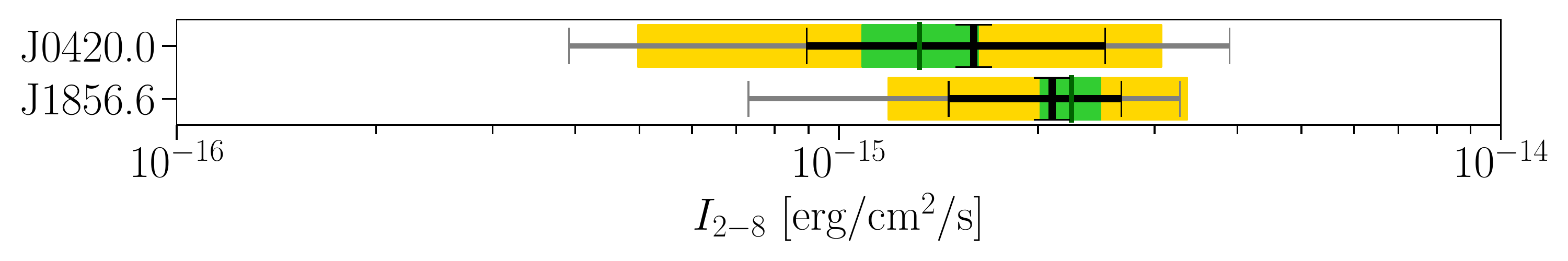}\\
\includegraphics[width=0.49\textwidth]{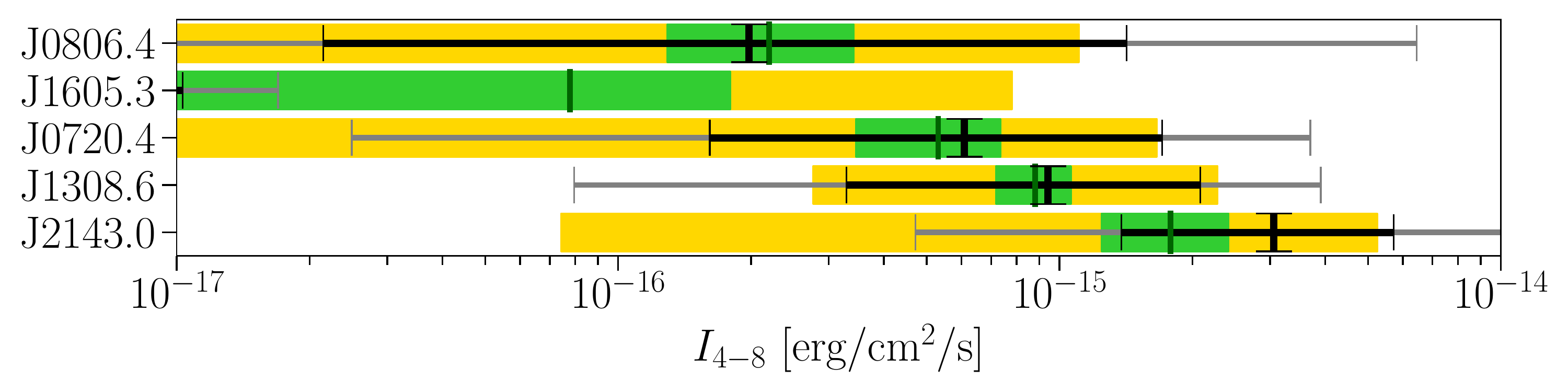}
\includegraphics[width=0.49\textwidth]{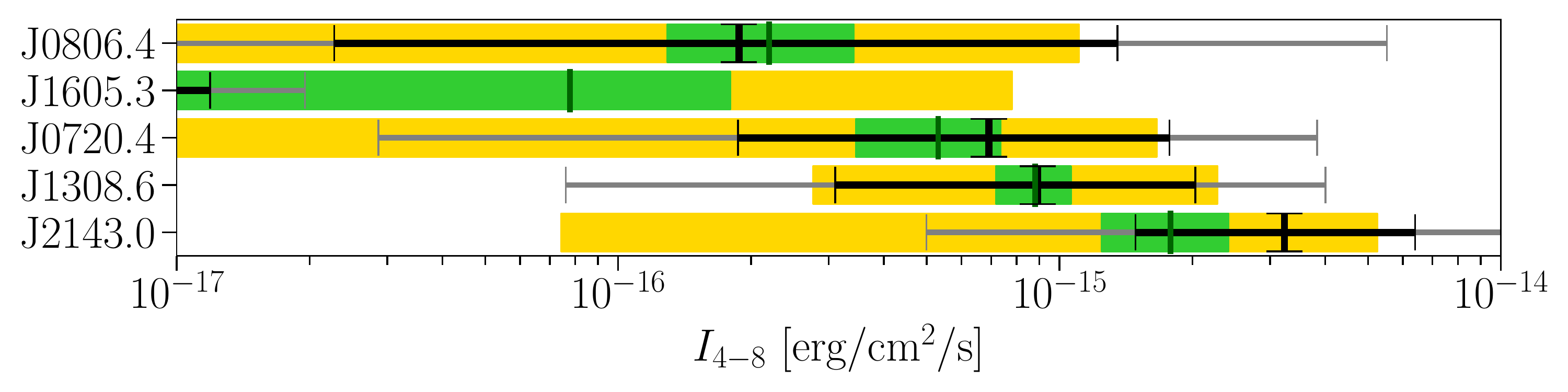}
\caption{
\label{fig:money-super}
As in Fig.~\ref{fig:money} and Fig.~\ref{fig:I28} but ({\it left}) assuming superfluidity model II and ({\it right}) assuming superfluidity model I.  Model II produces similar results to our fiducial analysis, which neglected superfluidity all together, while Model I leads to larger inferred axion couplings.
}
\end{center}
\end{figure}

The predicted $X$-ray flux depends sensitively on the assumed nucleon superfluidity model, since nucleon superfluidity suppresses the axion flux for temperatures below the critical temperature.  \new{Ref.~\cite{Potekhin:2020ttj} suggests that the nucleon superfluidity critical temperatures are likely too low to affect this analysis, and so we neglected superfluidity in the main body.  However, in this section we further illustrate the possible effect of nucleon superfluidity by considering a few superfluidity models in more detail. In our superfluidity model I~\cite{Page:2004fy}, we consider pure neutron pairing. In this subsection, we also consider two alternate models. Here, we do not consider the highly model-dependent PBF processes that we have explored in the previous section. In our alternative superfluidity model II~\cite{Elgaroy:1996mg}, we take into account that in a NS the neutrons are in $\beta$-equilibrium with the protons. This reduces the neutron effective mass and therefore reduces the range of densities in which the superfluid is allowed to form as well as the critical temperature at fixed Fermi momenta. Then the nucleon bremsstrahlung process is less suppressed, strengthening the limits as seen in Fig.~\ref{fig:money-super}. These models do neglect contributions from spin-orbit interactions, which have not yet been worked out but may prohibit superfluidity altogether~\cite{Schwenk_2004}.
}

Note that the critical temperatures in model II tend to be smaller than those in model I, and as a result the flux we obtain using model II receives a smaller superfluid suppression.
As seen in Fig.~\ref{fig:money-super} and comparing to Fig.~\ref{fig:money}, the superfluidity models make a significant impact on the best-fit axion parameter space.  Without nucleon superfluidity, and in the superfluidity model II, the best-fit axion couplings are significantly smaller than in model I since the axion production rates are highly suppressed in model I.

We note that since the critical temperatures depend on the Fermi momenta, which are determined by the EOS, uncertainties in the EOS also likely play a significant role in determining the uncertainties on the axion flux.  These uncertainties should be more thoroughly investigated in future work.

\subsection{Alternate magnetic field strength and core temperature models}

\begin{table}[]
\begin{tabular}{|c||c|c|c|c|}
\hline
Name  & $B_0$ [$10^{13}$ G]     & T$_{b}^\infty$ [keV]  & Refs. \\ \hline
J0806 & $9.2$    & ---                   & \cite{Haberl:2004xe} \\ \hline 
J1856 & $0.35$    & $7.8 \pm 3.9$       & \cite{Ho:2006uk,2013MNRAS.429.3517M} \\ \hline 
J0420 & $6.6$    & ---                   & \cite{Haberl:2004xe} \\ \hline 
J1308 & $4.2$    & $6.8 \pm 3.4$       & \cite{Schwope:2006ra,2011AA...534A..74H} \\ \hline 
J0720 & $3.0$    & $7.1 \pm 3.6$ & \cite{Hambaryan:2017wvm,2011MNRAS.417..617T}        \\ \hline  
J1605 & $1.0$    & $7.4 \pm 3.9$       & \cite{vanKerkwijk:2004qg,2012PASA...29...98T} \\ \hline 
J2143 & $14$    & ---                   & \cite{Kaplan:2009au} \\ \hline 
\end{tabular}
\caption{The alternate values of the magnetic field strengths and core temperatures, which are inferred from the kinematic ages of the NSs. The magnetic fields are derived from proton cyclotron absorption or NS atmosphere spectral fitting, and are given uncertainties by profiling over $\theta$.}
\label{tab:altNS}
\end{table}

In the main text we determined the core temperatures by extracting surface temperatures from a single blackbody fit to the 0.5 - 1 keV data (see~\cite{dessert2019hard}) and converting these temperatures to core temperatures as described in Sec.~\ref{sec:NS_Tb}. Here we investigate how the results change when we use the core-temperature estimates based solely on the kinematic ages of the NSs.  The core temperatures are inferred through the kinematic ages through the relation given in the main text.  Note that we assign a 50\% systematic uncertainty, translated appropriately to log space, on that relation to account for the precision quoted in~\cite{Beloborodov:2016mmx}.  We combine that systematic uncertainty with the uncertainties on the kinematic ages to produce the uncertainties quoted in Tab.~\ref{tab:altNS}.  We only include NSs in this analysis for which kinematic ages are known.

In the main text we adopted the magnetic field values determined by the spin-down rate of the NSs, except in the case of J1605 which has no measured value. These determinations give very precise measurements of the dipole component of the field, but the true field may be, {\it e.g.}, non-axisymmetric such that the spin-down measurements underestimate the magnetic field at the surface~\cite{Braithwaite:2008aw}. In this section we reanalyze the data assuming the magnetic fields determined by spectral fitting of the NSs while keeping the dipole assumption. These fields are inferred from cyclotron resonance absorption lines or NS atmosphere models. The fields are typically larger than the spin-down fields, although they are also significantly more uncertain, especially considering systematics such as the NS atmosphere composition. These alternate values are listed in Tab.~\ref{tab:altNS}.

\begin{figure}[htb]
\hspace{0pt}
\vspace{-0in}
\begin{center}
\includegraphics[width=0.49\textwidth]{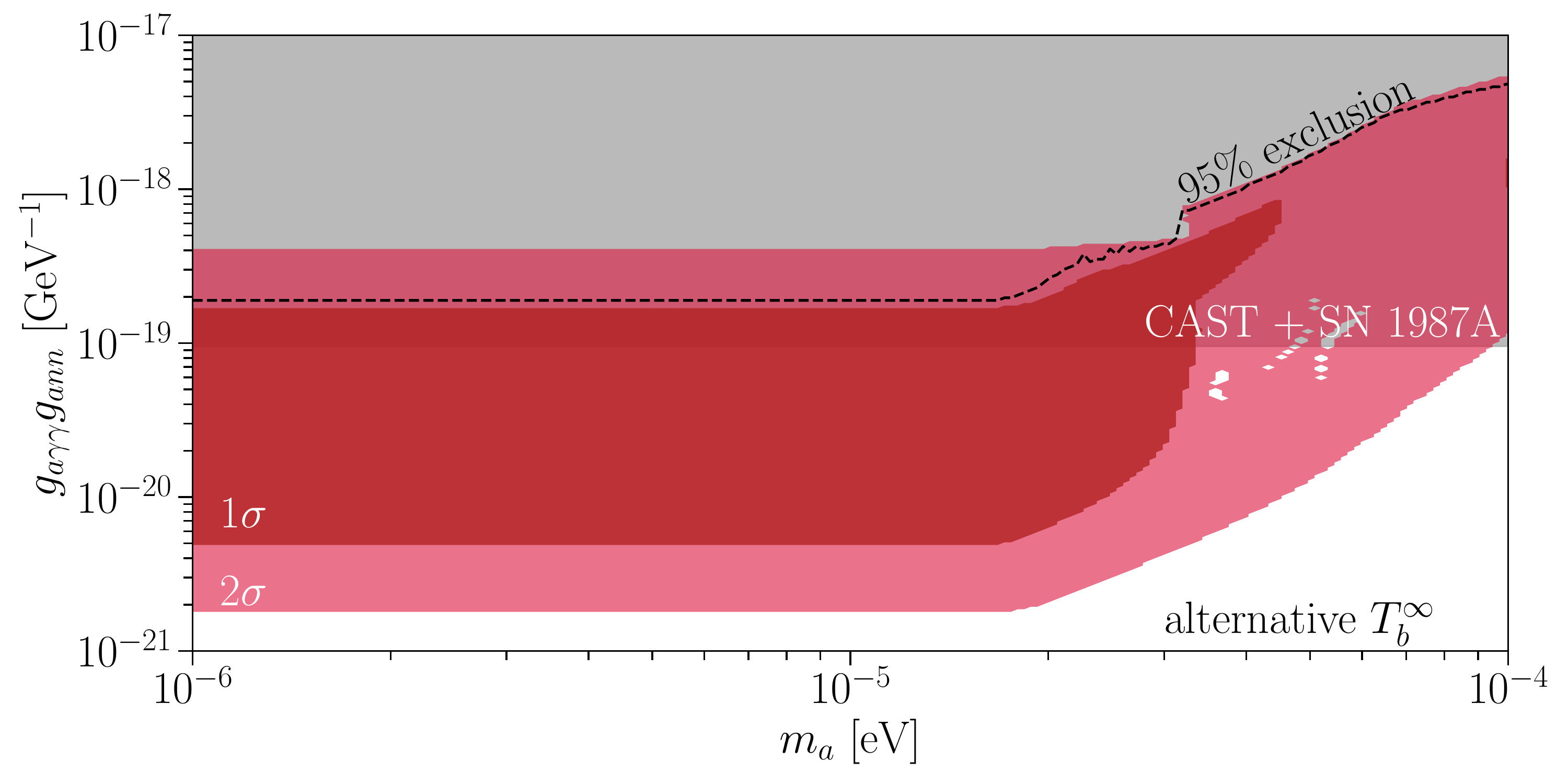}
\includegraphics[width=0.49\textwidth]{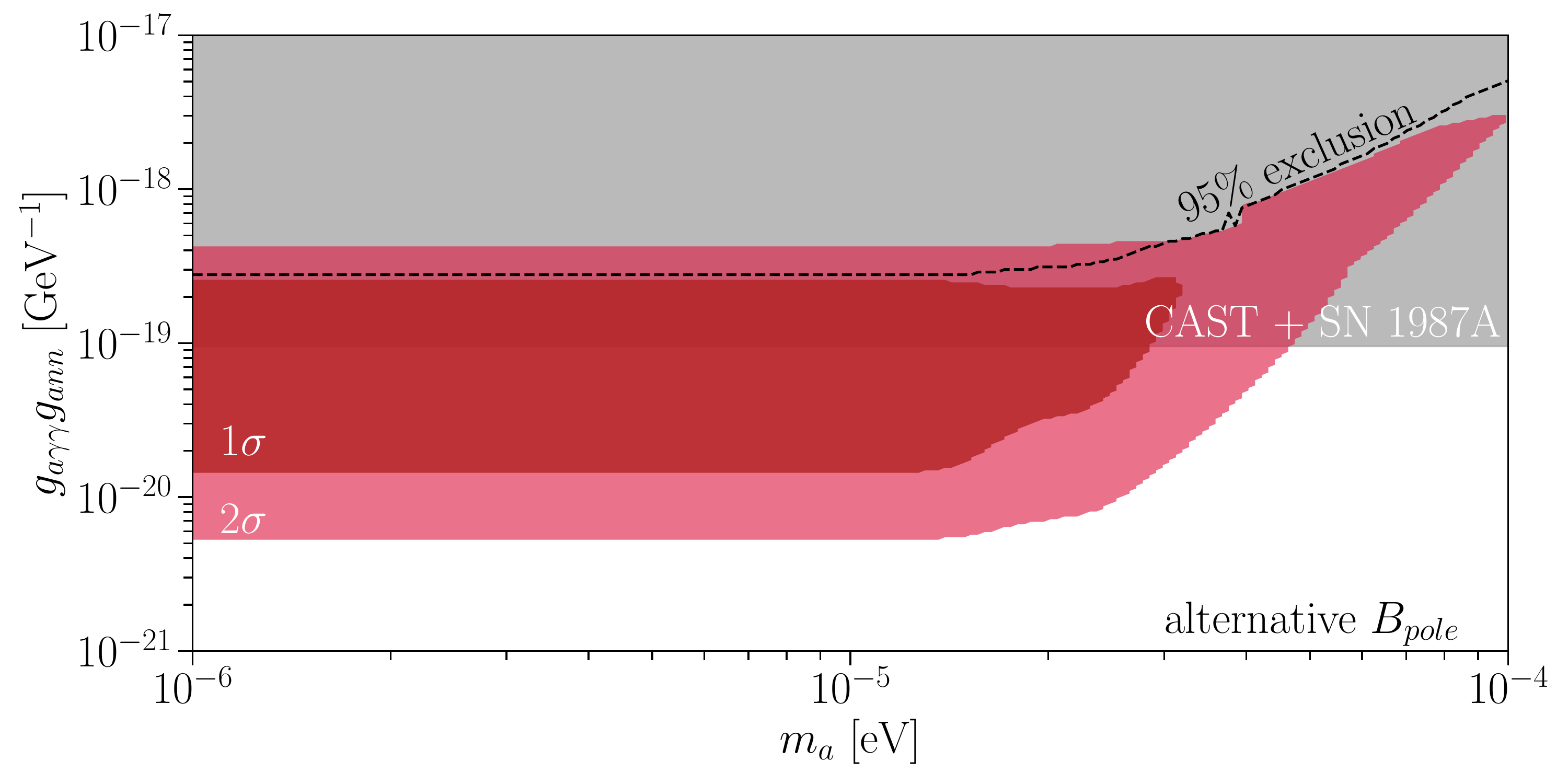}
\includegraphics[width=0.49\textwidth]{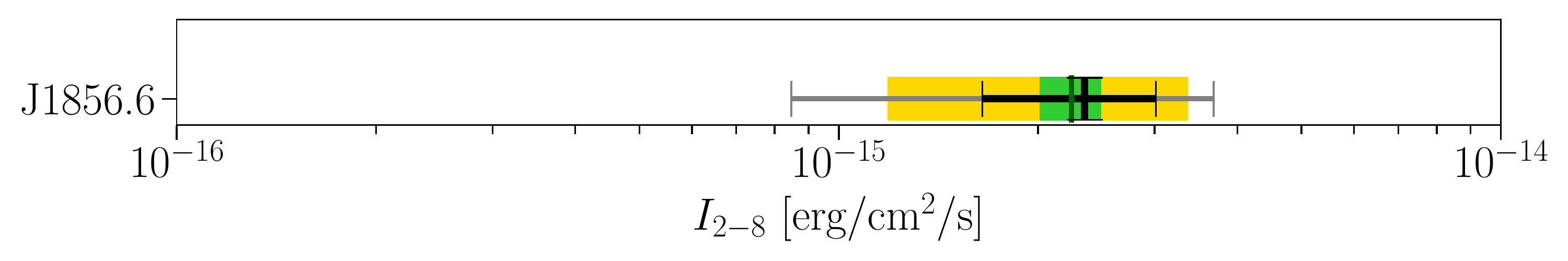}
\includegraphics[width=0.49\textwidth]{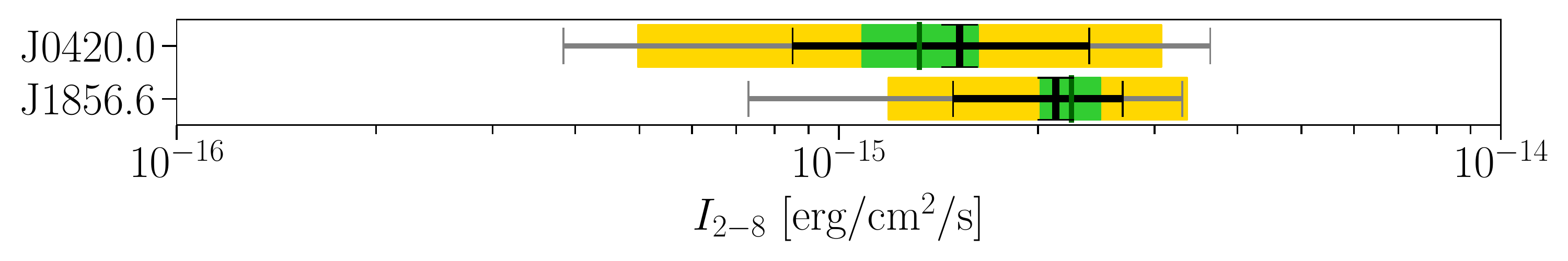}
\includegraphics[width=0.49\textwidth]{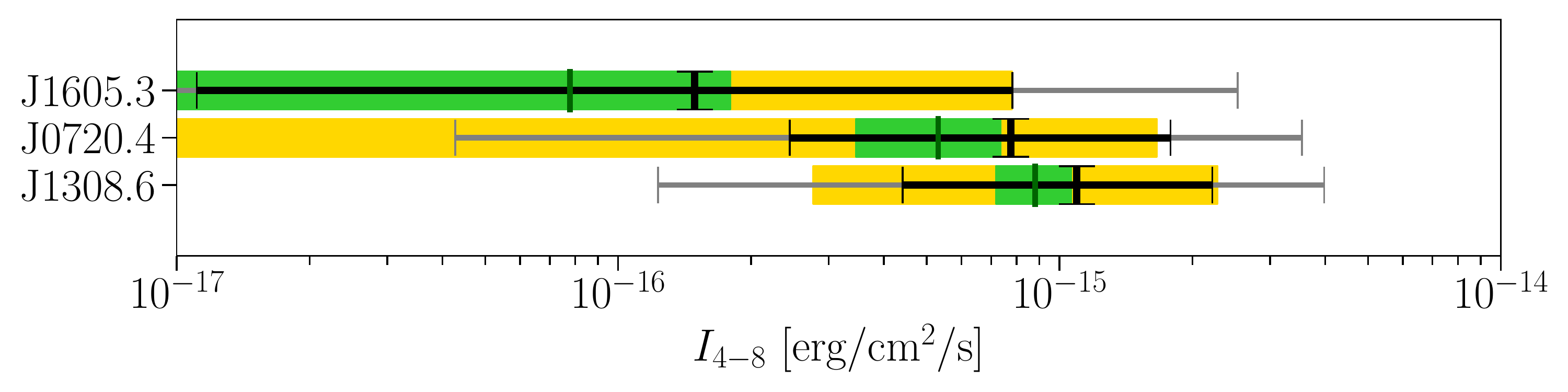}
\includegraphics[width=0.49\textwidth]{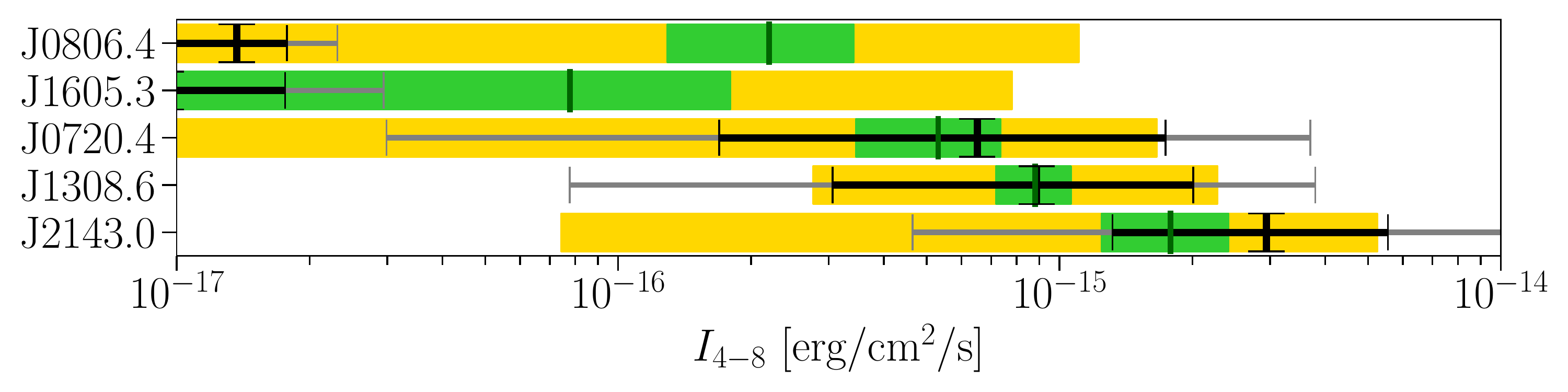}
\includegraphics[width=0.49\textwidth]{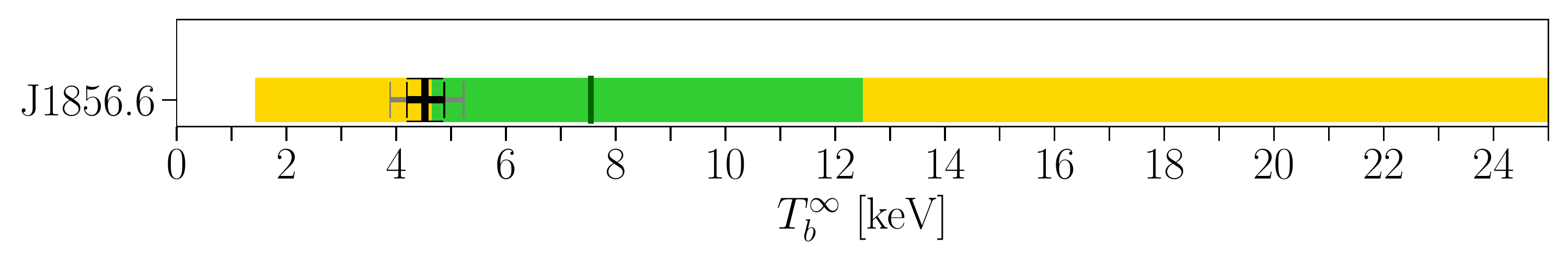}
\includegraphics[width=0.49\textwidth]{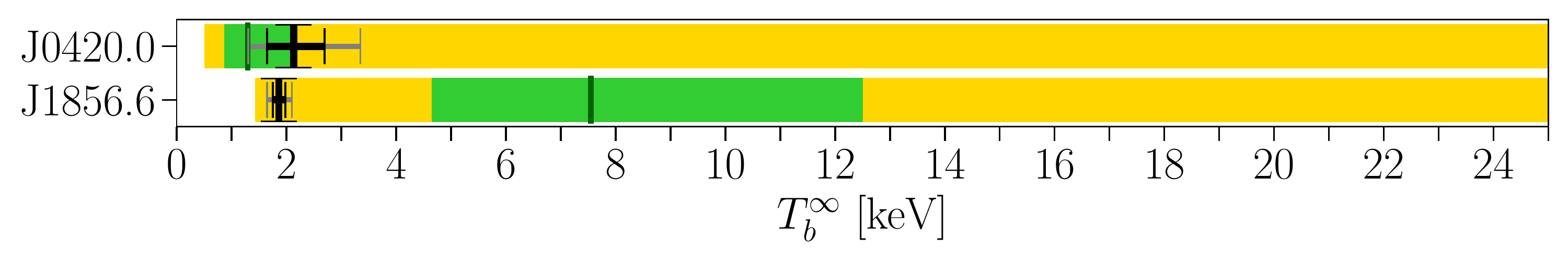}
\caption{
\label{fig:money-alt}
As in Fig.~\ref{fig:money}, Fig.~\ref{fig:I28}, and Fig.~\ref{fig:Ts} but ({\it left}) for the alternate core temperature values given in Tab.~\ref{tab:altNS} and ({\it right}) for the alternate magnetic field values shown in Tab.~\ref{tab:altNS}.
}
\end{center}
\end{figure}

In Fig.~\ref{fig:money-alt} we repeat our fiducial analysis ({\it left}) using the alternate core temperatures and ({\it right}) using the alternate magnetic fields.  The alternate magnetic fields have a relatively minimal impact on the best-fit parameter space. The alternate temperature model, on the other hand, has a more significant impact.  In this case the best-fit parameter space is at slightly higher axion couplings, due to the slightly lower core temperatures.  As seen in Fig.~\ref{fig:money-alt}, the alternate core temperature model also provides slightly improved consistency between the $I_{2-8}$ intensity observed from  J1856  and the $I_{4-8}$ intensities observed from the other three NSs considered, though we stress that this is a relatively minor difference.

\end{document}